\documentclass{ws-ijmpa}

\usepackage{xcolor}
\usepackage[verbose,hypertexnames=false]{hyperref}
\hypersetup{colorlinks=false,allbordercolors=blue,pdfborderstyle={/S/U/W 1}}
\usepackage[super]{cite}
\usepackage{url}
\usepackage{ulem}
\usepackage{paralist}

\newcommand{\bal}{\begin{align}}
\newcommand{\eal}{\end{align}}
\newcommand{\beq}{\begin{eqnarray}}
\newcommand{\eeq}{\end{eqnarray}}
\newcommand{\nneeq}{\nonumber \end{eqnarray}}

\newcommand{\nn}{\nonumber \\}
\newcommand{\es}{& = &}

\newcommand{\ps}{& + &}

\newcommand{\np}{\nn \ps}

\newcommand{\ket}[1]{ {\left|{#1}\right\rangle} }
\newcommand{\bra}[1]{ {\left\langle{#1}\right|} }

\newcommand{\bmat}{\left[\begin{array}}
\newcommand{\emat}{\end{array}\right]}

\newcommand{\ketbrac}[2] { {\text{$\mathfrak{C}_{{#1}{#2}}$}} }
\newcommand{\projector}[2]{{\text{$\mathbb{P}^{({#1})}_{{#2}}$}}}
\newcommand{\set}[1]{{\text{$\mathfrak{s}^{\dagger}_{{#1}}$}}}
\newcommand{\scrap}[1]{{\text{$\mathfrak{s}_{{#1}}$}}}
\newcommand{\qset}[1]{{\text{$\mathfrak{s}^{\dagger}_{{#1}}$}}}
\newcommand{\qscrap}[1]{{\text{$\mathfrak{s}_{{#1}}$}}}

\newcommand{\stepasym}[1] {{\text{$\mathcal{A}_{{#1}\leftarrow{#1}-1} $}}}

\newcommand{\innerid} {{\text{$\mathfrak{i}$}}}
\newcommand{\outeridentity} {{\text{$\mathbb{I}$}}}

\begin{document}
\markboth{J. J. G\'alvez-Viruet, Felipe J. Llanes-Estrada and Maria G\'omez-Rocha}{Preparations for Quantum Computing in Hadron Physics}

%
\catchline{}{}{}{}{}
%
\title{Preparations for Quantum Computing in Hadron Physics}

\author{J. J. G\'alvez-Viruet  and Felipe J. Llanes-Estrada}

\address{Theoretical Physics Dept. \& IPARCOS, Univ. Complutense de Madrid, Plaza de las Ciencias 1\\
28040 Madrid, Spain \\
juagalve@ucm.es}

\author{Mar\'{\i}a G\'omez-Rocha}
\address{Dept. de F\'{\i}sica At\'omica, Molecular y Nuclear
and Instituto Carlos I de F\'{\i}sica Te\'orica y Computacional,
Universidad de Granada, 18071 Granada, Spain}
\maketitle


\begin{abstract}
Quantum computers are coming online and will quickly impact hadron physics once certain
fidelity, decoherence and memory thresholds are met, quite possibly within a decade.
We review a selected number of topics where ab-initio Quantum Chromodynamics-level information about hadrons can be obtained with this computational tool that is hard to 
come by from other methods. This includes high baryon-density systems such as neutron-star matter (with a sign problem in lattice gauge theory); fragmentation functions; Monte Carlo generation of particles which accounts for quantum correlations in the final state; entropy production in jets; and generally, any application where time evolution in Minkowski space (as opposed to a Euclidean formulation) or where large chemical potentials play an important dynamical role. For other problems, such as the prediction of very highly excited hadron spectroscopy, they will not be a unique, but a complementary tool.
\end{abstract}


\newpage
\tableofcontents
\newpage
\section{Introduction: Strokebrushes of Hadron Physics {\it circa} 2025}	

If we had to put a date to the birth of hadron physics, it would probably be 1951, 
when Anderson, Fermi, Long and Nagle were for the first time artificially producing
the $\Delta$ resonance from a pion beam~\cite{anderson_total_1952}. 
Three quarters of a century have elapsed, and hadron physics has seen resounding successes and advances. The field is now immense, so the following lines provide only a few brushstrokes without capturing all its detail richness.

The hadron spectrum has grown to hundreds of particles and resonances by now. The basic mesons and baryons are all reasonably predicted by Monte Carlo methods in the Lattice Gauge Theory (LGT) formulation, after forty years of improvements~\cite{fukugita_hadron_1986,lang_vector_2015,durr_ab_2008}. The current thrust of experiment and phenomenology is in sorting out the excited spectrum above strong-force decay thresholds. This includes possible tetra/pentaquark states, also among them multiheavy hadrons (containing several heavy quarks)~\cite{celiberto_exotic_2024,alonso-valero_hartree-fock_2024}.  
LGT practitioners~\cite{Prelovsek:2025vbr} are also developing techniques to confront this excited spectrum, learning how to embed states in a continuum of several ground-state hadrons, and because such difficulties mostly stem from the discretisation of spacetime, {\it any} computer calculation, including those in quantum computers, will also have to face them. 

The same applies to the Lattice's difficulties with chiral fermions: the Nielsen-Ninomiya theorem~\cite{Nielsen:1981hk} applies equally to LGT whether formulated in Euclidean space for a classical computer or on Minkowski space for a quantum one. Thus, the quantum hardware with spacetime discretisation is not expected to be advantageous to deal with problems related to chiral symmetry and its breaking, such as the insensitivity (or sometimes ``restoration'', a less appropriate term) of the higher spectrum to chiral symmetry breaking~\cite{Bicudo:2009cr} bringing about parity doublets. 
In section~\ref{sec:particleregisters} below we discuss a particle--based encoding for quantum computers which might avoid some of these problems.

The one important question which  really remains unanswered in the hadron spectrum is Nathan Isgur's {\it Where is the glue\cite{Isgur:1996ke}?} Although hybrid mesons~\cite{Meyer:2015eta} and glueballs~\cite{Llanes-Estrada:2021evz}
have long been computed in LGT~\cite{Bernard:2003jd,Ape:1987thf,Athenodorou:2020ani}, their experimental extraction is, to say it candidly, stuck. Glueballs are mostly mixed with quarkonia into the meson spectrum; $J^{PC}$ exotics which could be hybrid-meson candidates do not abound, and the only broadly accepted one~\cite{ParticleDataGroup:2024cfk}, the $1^{-+}$ $\pi_1(1600)$ is somewhat too light to perfectly fit the bill as the expected exotic hybrid. In any case, it is unclear what contribution quantum computers could make here either: conventional theoretical techniques give reasonable estimates of the parameters and decay channels of hybrid mesons, with incremental but sure improvement. This does not mean that quantum computers should not be deployed to calculate the hadron spectrum~\cite{Gallimore:2022hai,deArenaza:2024dhe}, and indeed heavy mesons and all-heavy baryons in the Cornell model have already been recalculated employing the quantum computer as a small diagonaliser, but this should be seen as a means to calibrate the quality of their computations, not for the product of these.

\begin{figure}
    \centering
    \includegraphics[width=0.75\linewidth]{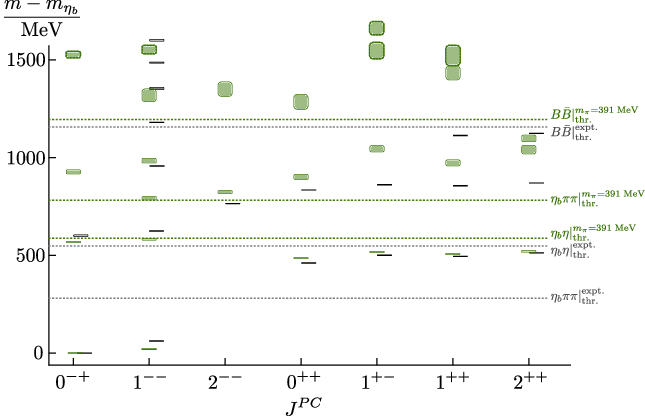}
    \caption{Bottomonium meson spectrum computed with Lattice Gauge Theory and its comparison with experimental data.
    {\it Reprinted~\cite{Ryan:2020iog} under the terms of the Creative Commons 4.0 License,} {\tt https://creativecommons.org/licenses/by/4.0/}
     }
    \label{fig:bb}
\end{figure}

An example of the immense progress of LGT in describing hadron spectroscopy is shown in figure~\ref{fig:bb}, displaying the bottomonium ($b\overline{b}$ mesons) computed on the lattice~\cite{Ryan:2020iog} and compared to the experimental status.

We can see that lattice computations can now access up to four excitations in a given $J^{PC}$ channel, at least for this relatively easy system (less so above the open--flavour thresholds).
Quantum computing can bring a different set of systematics to bear on the problem, as instead of operating computing correlators and looking for exponential fall-offs $e^{-Et}$ it can employ a systematic variational method with an increasing multiparticle Fock space to probe the eigenvalues of the Hamiltonian. So it can be deployed in the future as a check and in a complementary usage to LGT, though not as the primary means of computation, certainly not in terms of cost efficiency, for a foreseeable future.

Having discussed the modest interest of employing quantum computers in hadron spectroscopy, let us examine their possible role in advancing hadron structure and reactions.
Elastic and few-channel inelastic hadron-hadron interactions are well understood at low energies by a combination of methods: chiral effective theory~\cite{RodriguezEntem:2020jgp} and dispersion relations~\cite{Pelaez:2024uav}. These are clever parametrisations of the experimental data that allow some level of control of the systematic uncertainties~\cite{Salas-Bernardez:2020hua}, but have more parameter freedom than the very stringent number of parameters in the Lagrangian of QCD allows. 

Likewise, elastic or transition form factors~\cite{Lorenz:2012tm,Raya:2015gva}
are informative about the properties of hadrons when they actually behave as hadrons or at most display a few constituent-like degrees of freedom, and 
LGT~\cite{Barone:2025rye} does make steady progress towards their description with full uncertainty budgets from extrapolating to continuum grids, infinite volume and quark masses at the physical point, because the relevant matrix elements can be analytically continued to Euclidean regions of the momentum variable where the statistical formulation in a lattice presents no fundamental problems.

Deeply inelastic scattering and other processes involving a high-$Q^2$ probe of a hadron entail more difficulties to traditional LGT. Progress has been made in combining a large-momentum effective theory, LAMET, with lattice techniques~\cite{Good:2025daz}, and it appears that parton distribution functions, quite well measured and constrained by experimental data, are starting to be amenable to calculation.

Next, let us briefly discuss the collective medium formed in heavy ion collisions. A new phase of matter has been found in these, which behaves as a fluid~\cite{Shuryak:2003xe}, not a rarefied gas of particles. Near the phase transition to a more conventional hadron gas, it has low viscosity~\cite{Romatschke:2007mq}, relative to its entropy density. We hesitate to call it, as is common, a ``Quark--Gluon Plasma''~\cite{Shuryak:1977ut} because of this strongly-coupled fluid behaviour as well as the predicted appearance of chains or loops of confined colour charges, which would make its interpretation as a plasma very difficult, 
so a ``Quark--Gluon Gasoline'' would be a closer analogy  at the temperatures of order 150 MeV near the phase transition~\cite{Fujimoto:2025sxx,Cohen:2024ffx}. Be it as it may, Lattice Gauge Theory~\cite{Philipsen:2024hla} is a tool which can calculate many properties of hadron matter at very large temperatures, as long as the baryon chemical potential is negligible (see Section~\ref{sec:openproblems} shortly). 

Many more interesting topics have developed within hadron physics in the last years. For example, the so called ``Strong CP problem'' seems to have been put to rest~\cite{Ai:2020ptm,Ai:2024cnp} by showing that the $\theta$ parameter does not show in CP-violating observables, if the limits of large volume and sum over topological sectors are taken in this order, unlike in earlier work which produced a CP-violating phase from the $\theta$-vacuum. This is the realm of analytical work.

Also, artificial intelligence is making inroads in our field\cite{Vent:2025ddm,Sadasivan:2025kjj}, to assist event or jet classification, or identifying parameters of scattering amplitudes such as pole positions, for example. All these examples, and many more, show the richness of physics and methods in hadron physics: many of them will not see substantial progress due to the arrival of quantum computers. However, there are a few interesting open problems where they {\it will} have an impact, and we devote the next section~\ref{sec:openproblems} to list three of them. 
It is our intent in this work to confine ourselves to topics in which individual hadrons, their excitations, and their quark-gluon components come to the fore, avoiding excursions into either High Energy Physics or Nuclear Physics, where other reviews have already appeared. Additional material, with overlaps to nuclear physics and to particle physics, can be found, for example, in the white paper~\cite{Beck:2023xhh} for the US DOE.

\section{Some niche problems which remain unaddressed with first principles}
\label{sec:openproblems}
In this subsection we briefly explain the difficulty in computing, from first principles,
certain niche problems with conventional methods in hadron physics. Later on we will dedicate some paragraphs to explain what progress has been made and can be expected from the quantum computing approach.

\subsection{Equation of state of neutron stars}
The behaviour of an atomic nucleus composed of a few nucleons~\cite{Shen:2024qzi} is well characterised by chiral interactions. Nuclear matter below the saturation density $n_s$ may be addressed with the same chiral interactions or even directly from scattering data without assuming a Hamiltonian~\cite{Alarcon:2022vtn}.
But as density increases, probably around 1.5-2 times $n_s$, certainly above 2-2.5 $n_s$, the chiral expansion breaks down and uncertainties become unmanageable~\cite{Tews:2024owl}.

\begin{figure}[h!]
    \centering
    \includegraphics[width=0.7\linewidth]{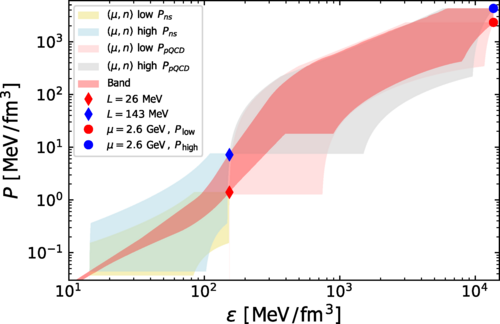}
    \caption{Status of the Equation of State of Neutron Star  Matter~\cite{Alarcon:2024hlj,LopeOter:2025fbt}. To the left, low--density side, chiral interactions, nucleon scattering and nuclear data are quite constraining. To the right, at asymptotically high densities (beyond the presumed range of neutron--star ones) pQCD gives a controlled uncertainty band. However, through a large swath of the diagram, only causality and stability (as well as integral constraints from the $n-\mu$ plane, shown in different shades) are available. It is at these densities, unreachable by Lattice techniques that a quantum computer could make a significant contribution. 
    {\it Copyright: American Physical Society. Reproduced with permission.}}
    \label{fig:EoS}
\end{figure}

From there on, the allowed band of Equations of State inside a neutron star is constrained only by fundamental theory, namely by causality, $c_s^2\leq c^2=1$ and mechanical stability, $c_s^2\geq 0$; since $c_s^2=\frac{dP}{d\varepsilon}$ with $P$ the pressure and $\varepsilon$ the mass-energy density. This is seen in figure~\ref{fig:EoS}.

More theoretical  information is available at asymptotically high densities where QCD becomes weakly coupled and the Equation of State is calculable in a loop expansion~\cite{Fraga:2013qra}.
This is why the uncertainty band is seen to narrow towards the right of the figure leaving a ``Rhoades-Ruffini'' diamond~\cite{Rhoades:1974fn} (distorted here because of the logarithmic scales used).

Narrowing the band of possible equations of state in figure~\ref{fig:EoS} as much as possible is important to potentially run tests of the theory of gravity with good knowledge, from microscopic physics alone, of the stress-energy tensor, the right hand side of Einstein's field equations (for an isotropic and homogeneous fluid at rest, $T={\rm diag}(\varepsilon, P,P,P)$):
\begin{equation} \label{Einstein}
\frac{c^4}{8\pi G_{\rm N}} \  G_{\mu\nu} = T_{\mu\nu}\ .
\end{equation}
How could a quantum computer help here?

One possibility is to work from the nuclear side (from left to centre of the figure) by employing the quantum computer, basically, as a diagonaliser, profiting from its large (variational) Hilbert space, with basis of size $N=2^{n_q}$ exponentiating the number of qubits $n_q$. This method would not suffer from the breakdown of the perturbative expansion in Chiral Perturbation Theory upon computing thermodynamic properties. However it would still be subject to the failure of the theory in providing an {\it ab initio} Hamiltonian as its organising principle, that Chiral Expansion in powers of momentum etc. would fail; even the degrees of freedom, nucleons, would be in doubt at such large densities that the vacuum symmetries~\cite{Llanes-Estrada:2011nal,Buballa:2015awa} (from which the Lagrangian is constructed) could change.

A probably more rewarding procedure consists in employing the quantum computer together with a quark--level Lagrangian, and try to calculate the EoS from high to low densities (right to left in figure~\ref{fig:EoS}).  We will dedicate Section~\ref{sec:eos} below to describing
an ongoing effort in this direction.

\subsection{Fragmentation functions in hadronising jets}
\label{subsec:frag}

\begin{figure}
    \centering
    \includegraphics[width=0.75\linewidth]{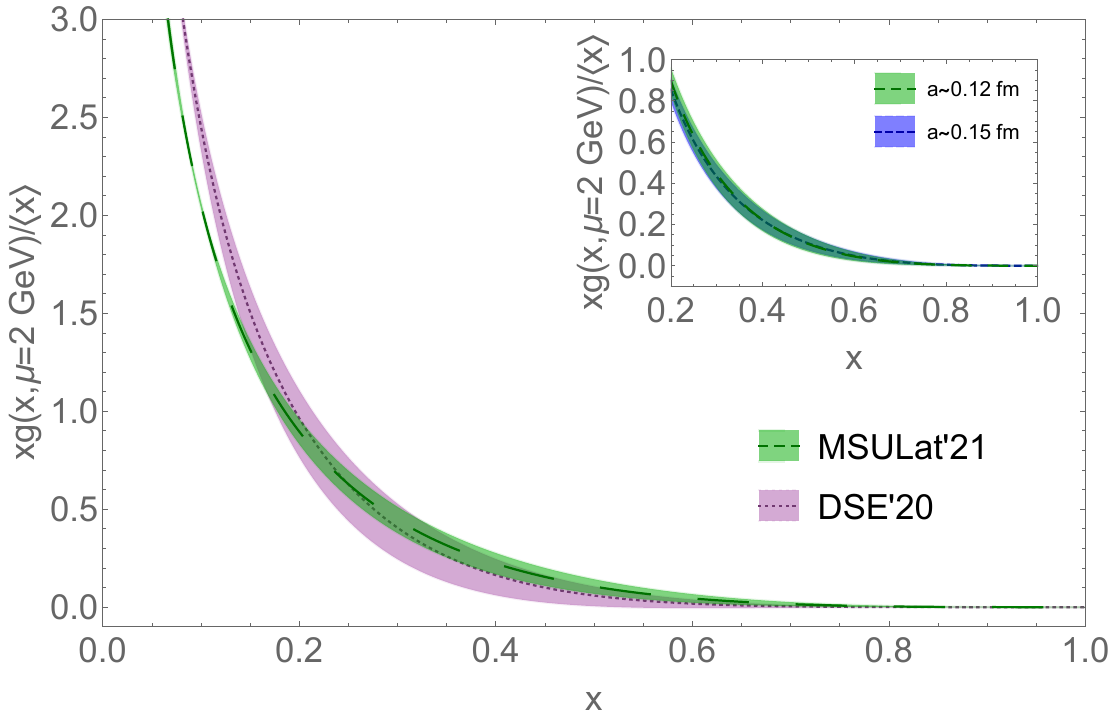}
    \caption{The gluon parton distribution function $xg(x)$ inside the kaon computed from LGT compares very well with Dyson-Schwinger estimates and experimental data--driven extractions above $x\geq 0.2$ and seems qualitatively correct over the entire $x$ range. {\it Reprinted from~\cite{Lin:2025hka} with publisher's permission.}}
    \label{fig:gluonpdf}
\end{figure}

A large historic effort in Lattice Gauge Theory has finally provided some substantial progress in computing parton distribution functions $f(x)$ {\it ab initio}, first by computing their ``momenta'' (projections of $f(x)$ with appropriate weights, $\int_0^1 f(x) x^n dx$) and more recently by applying Effective Field Theory methods which allow better access to the Bjorken--$x$ dependence~\cite{Lin:2025hka}.
Serve as example the gluon distribution function inside the kaon shown in figure~\ref{fig:gluonpdf}.

Not such progress has been made, however, in computing fragmentation functions $D(x)$.
These are nonetheless extremely interesting objects because {\it (a)} they grant access to all hadrons which can be reconstructed in the final state, and not only the much smaller number of them which are stable enough to be prepared for the experiment in the initial state; and {\it (b)} they must be sensitive to how quarks and gluons confine into hadrons, the asymptotic end states in any hadron process (a suggestive image for this jet hadronisation is to think of a nut, bolt or spring moving at large speed, picking up other pieces as it propagates, and producing an entire mechanical clock in the end state). We do not have so many experimental windows to actual confinement mechanisms such as the centre-vortex one, which remains mostly a theoretical endeavour.

A possible definition of fragmentation functions employs the probability of finding a hadron $h$ plus other particles $X_{\text{out}}$ not reconstructed,  within the jet that was launched by a given bare quark or gluon $j$ whose longitudinal momentum fraction $z=p_h^-/p_q^-$ is the argument of the fragmentation function~\cite{Gronau:1973gc,Feynman:1973xc,Collins:2023cuo} $D^h_{j}(z)$, which at given renormalisation scale and on a discretised momentum space is defined by
\begin{eqnarray}
    D_{j}^h(z) 
    \equiv \frac{\rm Tr_{c} } {N_{c,j}}\sum_X 
    \langle j,p_1 | h,X_{\text{out}}\rangle 
    \langle h,X_{\text{out}} | j,p_1\rangle,
     \label{FuncionFragmentacion}
\end{eqnarray}
with $N_{c,j} $ being the number of colours, and the trace also being applied over parton colour space.

Emission of a hadron from a parton $p$ originating in an $e^-e^+$ annihilation is, in terms of the fragmentation function $D$,
\begin{equation}
\frac{d\sigma|_{e^-e^+\to h(z)X}}{dz}\!\!  = \!\! 
    \sum_p\!\! \int_z^1\!\! \frac{dy}{y}  \frac{d\sigma|_{e^-e^+\to p(y)X}}{dy}D_a^h(z/y).
\end{equation}
but the fragmentation functions are universal: it is irrelevant that the parton came from a lepton-lepton reaction instead of a deep-inelastic scattering one on the nucleon or a purely hadronic one.
A fixed hadron emitted by a fixed parton requires convoluting the pQCD-computable kernel $C_l$ with the universal fragmentation function $D_q^h(z)$.

Currently, fragmentation functions are fit to data from  high-energy experiments~\cite{AbdulKhalek:2022laj,Bertone:2018ecm}; but they have proven unassailable by direct {\it ab initio} calculation.
The Euclidean space formulation of LGT requires time to be continued $t\to x_0=it$ to the imaginary axis. However, the field-theoretical equation extending Eq.~(\ref{FuncionFragmentacion}) which is appropriate for a spacetime lattice is~\cite{Collins:2023cuo}
\begin{eqnarray}
D_{ j}^h(z) = \frac{z^{1-2\varepsilon}\rm Tr_D}{4} \sum_X
\int \frac{dx^+}{2\pi} e^{ik^-x^+} \gamma^- \cdot \nonumber \\ 
\langle 0 | \psi_j\left( \frac{x}{2}\right) | h,X_{\text{out}}\rangle     \langle h,X_{\text{out}} | \bar{\psi}_j \left( -\frac{x}{2}\right) | 0\rangle \ .
\end{eqnarray}
Euclidean LGT stumbles because the fields are taken at different points along light--front time (note the integral over $x^+$), an evolution operator along a lightlike direction is needed (which is not possible in an Euclidean formulation with four equivalent spacelike directions). 

Now, near-to-medium future quantum computers promise quick improvement.
Currently, Nambu-Jona-Lasinio model computations   are already at hand~\cite{Li:2024nod}, and our group is actively exploring the hardware needed to proceed to appropriately regulated Quantum Chromodynamics, with its more fundamental Hamiltonian formulated in Light-Front quantisation (see subsec.~\ref{subsec:lightfront} and section~\ref{sec:fragmentation} below).

\subsection{Monte Carlo event generation}
Another interesting example which affects many experimental collaborations in high-energy
physics is particle production in Monte Carlo generators. The best known example
is {\tt PYTHIA} in which from the initial hard partons, strings with quark/antiquark sources
are generated that then produce the final state hadrons. This production process follows
a classical probability distribution. It has been very successful, for decades now, 
in giving a rough parametrisation of one--body spectra (with various parameter ``tunes'')
and is widely appreciated by its versatility and speed in producing numerous multiparticle
events.

\begin{figure}[h!]
    \centering
    \includegraphics[width=0.32\linewidth]{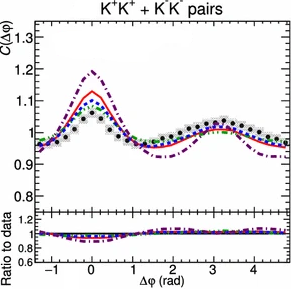}\ 
    \includegraphics[width=0.32\linewidth]{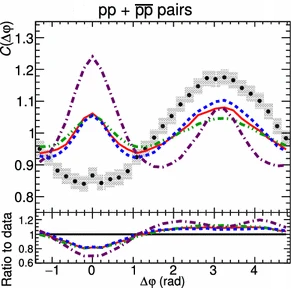}\ 
    \includegraphics[width=0.32\linewidth]{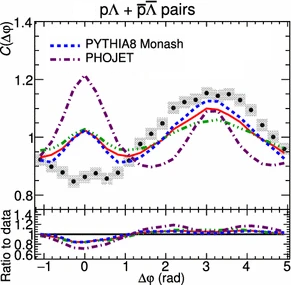}
    \caption{From left to right: a)  Typical ALICE~\cite{ALICE:2016jjg} experimental data and Monte Carlo simulation of two-meson correlation against $\varphi$, with a peak at $\phi_1-\phi_2=0$ indicating that the mesons are positively correlated. b) and c) In the experiment, the baryon-baryon and antibaryon-antibaryon correlations dip below 1, implying anti-correlation.
    In this baryon case, Monte Carlo simulations disagree with the experimental data. This is true for both identical and different baryon species.
    Copied from~\cite{ALICE:2016jjg} by the ALICE collaboration under the terms of the Creative Commons License 4.0 (http://creativecommons.org/licenses/by/4.0/); no changes have been effected.}
    \label{fig:Baryoncors}
\end{figure}

However, in the last years, the flourishing experimental works on particle-particle two-body correlations have stressed the program. For example, the ALICE collaboration~\cite{ALICE:2016jjg}
reported a qualitative disagreement between their data, shown in figure~\ref{fig:Baryoncors} 
and that shows baryon-baryon {\it anti-correlation}  (if a baryon is emitted in a given azimuthal direction, it is less likely than a second baryon is emitted along a nearby direction), unlike mesons whereas the Monte Carlo simulation yields a positive correlation analogous to that of mesons (which are bosons and are likely to be closely emitted in momentum space). 

A possible ad-hoc fix~\cite{Demazure:2022gfl} (see~\cite{Lonnblad:2023kft} for further discussion) 
exemplifying the root of the problem
in the Monte Carlo simulation, is to spread the baryons in phase space, with two rules
``One-baryon policy'' and ``All baryon policy'' which force every string to produce
one and only one baryon (the first rule to guarantee the anti-correlation, the second
to avoid lessening the overall number of baryons). 

We can also conjecture, although we are not aware of an investigation in this respect, than in circumstances in which experimental data may violate a Bell inequality, the simulation will however satisfy it, because its probabilistic implementation is entirely classical. 

A much more satisfactory solution would be to have a full quantum calculation where the Pauli principle among quarks is satisfied and where correlations and anti-correlations satisfy the same quantum rules that apply in the real world. This is however computationally unfeasible due to the huge running times: a quantum computer, however, 
could probably speed-up the hadronisation steps and produce more natural (anti)correlations among particles.

\section{Current state of and prospects for quantum computing}

\subsection{Digital versus Analog Simulations}
Digital simulations\cite{Georgescu:2013oza} of time evolution in a quantum system (such as a hadron) are based on the application of stroboscopic unitary evolution, first proposed by Lloyd~\cite{Lloyd:1996}, on universal quantum computers. 
If a Hamiltonian consists of the sum of several terms $H=\sum_j H_j$, a Suzuki-Trotter type formula with stroboscopic (slow-motion) steps is employed
\begin{equation}
   e^{-iHt} \simeq \left( 
   \prod_j e^{-iH_j \frac{t}{n}}
   \right)^n,
\end{equation}
where the error incurred in discarding the commutators $[H_j,H_{k\neq j}]$ is of order $(t/n)^2$ and systematically controllable, in principle, by increasing $n$.

In contrast, analogue quantum computations are based on the fact that there are systems, the simulators, that can be tuned to behave in certain regions of parameter space as another system which is in principle inaccessible\cite{Georgescu:2013oza}. The mapping should relate the states of both systems, if $\left|\phi_t\right\rangle$ is an state of the objective system and $\left|\psi_t\right\rangle$ the corresponding state on the simulator, there should exist a bijection such that 
\begin{equation}
f\left(\left|\phi_t\right\rangle\right) = \left|\psi_t\right\rangle,
\end{equation}
so that the Hamiltonians are related by
\begin{equation}
    H_{sim} = fH_{phys}f^{-1}.
\end{equation}
In general analogue simulators are robust against errors, at least more than their digital counterparts and there is no need of Trotter decompositions to implement unitary evolution.

Both digital and analogue quantum computers can be used to execute quantum algorithms. The work of Jordan, Lee, and Preskill\cite{Jordan:2012xnu} about real-time scattering in $\phi^4$ theory was one of the earliest algorithms in the context of high-energy physics. The simulation of a scattering event was structured into three stages: {\it a)} preparation of the asymptotic wavefunction, {\it b)} time evolution through the scattering region, and {\it c)} measurement of the outgoing states. These stages mirror the fundamental routines of every quantum computation: state preparation, unitary evolution, and measurement and can be executed in both types of computers. The asymptotic costs of each step were also calculated, demonstrating that the process can be efficiently simulated. Numerous other theoretical studies followed on Lattice Gauge Theories \cite{Byrnes:2005qx,Zohar:2014qma,Marcos:2014lda,Zohar:2015hwa,Lamm:2019bik,Mezzacapo:2015bra,Davoudi:2022xmb}, spectroscopy\cite{Gallimore:2022hai,deArenaza:2024dhe}, jet evolution\cite{Barata:2021yri,Barata:2023clv,Qian:2024gph}, fragmentation\cite{Li:2024nod}, encoding~\cite{Barata:2020jtq,Kirby:2021ajp,Kreshchuk:2023btr}, and more~\cite{DiMeglio:2023nsa}; and experimental tests, see for example \cite{Martinez:2016yna,Klco:2019evd,Atas:2022dqm,Farrell:2022wyt,Farrell:2022vyh,Farrell:2024fit,Roland:2025} for digital computations and \cite{Friedenauer:2008gny,Gerritsma:2010bpn,Zhang:2016lyo,Silva:2019,Gonzalez-Cuadra:2024xul} for analogue experiments. The reader is referred to the subsections below for recent applications to hadron physics.

\subsection{Hardware} \label{sec:hardware}
Impressive progress in quantum computing hardware, here briefly summarised, and its availability through cloud services for researchers worldwide, is causing an explosion of interest and many exploratory works have appeared and continue to appear.

Simulations on quantum computers are based on the application of successive unitary transformations $\hat{U}$ on $\textit{quantum memories}$. These quantum memories are most often composed of spin-$1/2$ systems which implement the qubits, so the total Hilbert space is $\mathcal{H} = \left(\mathcal{H}^{1/2}\right)^{\otimes n}$, where $n$ is the total number of qubits.

Since Feynman and Deutsch initial research on quantum computers \cite{Feynman:1982,Deutsch:1985vkq} there has been a continuous theoretical effort in precisely defining the characteristics and requirements of fault-tolerant computations\cite{DiVincenzo:2000tra} as well as many experimental efforts in different hardware platforms to realise those requirements \cite{Ladd:2010cup}, we here briefly mention some of these platforms: \textit{superconducting circuits}, \textit{trapped ions}, \textit{linear photonic circuits} and \textit{Rydberg atoms}.

\begin{figure}
    \centering
    \includegraphics[width=0.4\linewidth]{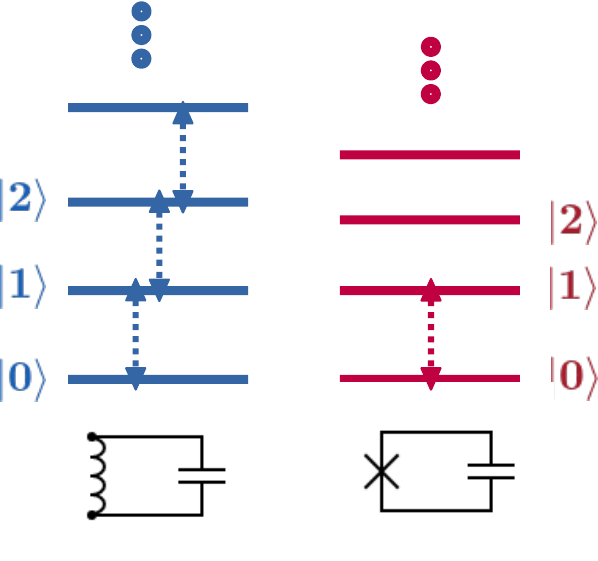}
    \caption{Left: Energy--level scheme of a quantised harmonic oscillator $H={\rm constant}+\hbar\omega(a+a^\dagger)$ (implementable as a microscopic oscillating $LC$ circuit). Microwaves which excite the $|0\rangle \leftrightarrow |1\rangle$ transition can also excite higher levels. Right: to implement a two--level ``transmon'' qubit, a nonanharmonicity is needed, which is achieved by swapping the linear inductor by a non-linear Josephson junction. The Hamiltonian receives nonharmonic terms such as $-\hbar \omega'(a+a^\dagger)^4$, and the equispacing is broken, so that no further excitation of the oscillator is possible with the same microwave pulse.}
    \label{fig:transmon}
\end{figure}

The currently most used of those platforms deploys \textit{superconducting circuits}, based on \textit{transmon qubits}~\cite{Wallraff:2004,Koch:2007hay,Krantz:2019jkw}. They combine a non-linear Josephson junction between two superconductors and a capacitor to form an anharmonic oscillator, see figure~\ref{fig:transmon}.
The resulting oscillator is sometimes dubbed an ``artificial atom''\cite{Georgescu:2013oza}, whose two lowest energy levels, uniquely linked by $E_1-E_0$ microwave photons which can excite no other transition,  provide the qubit states. Present day systems put together well past several hundreds of qubits, albeit with insufficient gate fidelities to implement error--correction protocols \cite{Arute:2019zxq,Zhu:2021gkn, IBMquantum:2025, IBMroad:2025}. Therefore, other alternatives are being explored in parallel. 

\textit{Trapped ions}\cite{Cirac:1995zz,Friedenauer:2008gny,Martinez:2016yna} is another promising technology for quantum computing and simulation. The basic building blocks are ion chains confined in optical traps. The ion's internal states provide the qubit's $|0\rangle $ and $|1\rangle$. Single-qubit transformations are then realised by the excitation of individual ions with laser pulses, while entanglement can be introduced by changing the collective motions of the particles in the trap. Remarkably, these technologies give access to long-range interactions and the possibility of \textit{qudit} computations~\cite{Ringbauer:2021lhi}, where the basic blocks are quantum systems with dimension higher than 2.  Companies such as {\tt IonQ} have commercially available systems with 25 and 36 qubits which can be accessed on the cloud; or {\tt Alpine Quantum Technologies} that sells an office machine with 20 qubits.
In addition to good two-qubit gate fidelity (the AQT system claims to benchmark at 0.9847(3)) and decoherence time, these platforms offer interconnectivity of all qubits (which avoids {\tt SWAP} operations to control remote qubits by exchanging their states to next-neighbour ones as in other architectures).

\textit{Linear photonic} circuits are another promising avenue for the development of universal quantum computation. Photons provide inherently coherent and nearly noiseless qubits, allowing single-qubit operations to be implemented with high fidelity\cite{Slussarenko:2019flf}. However, this same feature makes the realisation of multi-qubit gates particularly challenging. In fact, instead of relying on strong nonlinear interactions—required for direct multi-qubit operations—photonic quantum processors typically employ the so-called KLM scheme\cite{Knill:2001lrt}, where effective nonlinearities are induced through measurements, rendering entangling gates non-deterministic (presenting a finite failure probability). This is hardly acceptable for calculations, such as the ones described in this review and many others which will need millions to billions of logic gates. The probability of a flawless calculation would be negligible and the cost to obtain it would exponentiate with the number of entangling gates. 

Near-deterministic gates are therefore under implementation. They are meant to combine multiple parallel circuits with quantum teleportation protocols; the non-deterministic transformations are prepared on independent, "auxiliary" circuits and, upon success, their results teleported to the main qubits. This, one hopes, will avoid the exponential growth of resources needed for successful computation with large numbers of non-deterministic gates. 

The field is evolving rapidly, and several promising approaches are emerging, including the development by {\tt PsiQuantum} of a silicon-based, chip-like platform that integrates all necessary photonic components~\cite{PsiQ:2025}.

\textit{Neutral atom} circuits rely on arrays of neutral atoms confined in magnetic or optical traps, whose valence electrons are excited to high principal quantum number states, the so-called {\it Rydberg states}~\cite{RydReview:2021}. These states are characterised by large dipole moments and, being very peripheral, also large mean radii, leading to strong dipole–dipole interactions and to the Rydberg blockade phenomenon, which prevents two nearby atoms from being simultaneously excited to their Rydberg states. 

Single-qubit gates are typically implemented through laser-mediated Raman transitions, in close analogy with the trapped-ion quantum computers described two paragraphs above. Two-qubit gates, on the other hand, can be realised via dipole–dipole interactions when the interaction strength is small compared to the excitation strength. 

In the opposite regime—dominated by Rydberg blockade—an excited atom inhibits nearby atoms from accessing their Rydberg states, effectively serving as a control qubit. A key advantage of neutral-atom systems lies in the favourable ratio between the gate execution rate, set by the excitation strength to Rydberg states ($g \sim 10^{9}\text{ s}^{-1}$), and the dephasing rate of atoms in the trap ($\lambda \sim 0.1\text{ s}^{-1}$), which in principle allows for up to $g/\lambda \sim 10^{10}$ coherent gate operations. Nonetheless, these platforms currently face important limitations in multi-qubit gate fidelities—the highest demonstrated being $\mathcal{F} < 0.8$—while fault-tolerant quantum error correction requires $\mathcal{F} > 0.9999$~\cite{Saffman:2019pvh}.

In summary, quantum technologies are advancing rapidly and allowing the implementation of quantum computers that can already surpass their classical counterparts in specific tasks \cite{AbuGhanem:2023kpk}. 

Based on our group's work, we can state that while the number of qubits in the superconducting platforms is now starting to be satisfactory for not too demanding applications, the number of consecutive gates that can be executed without being overrun by noise has to increase.

For the extraction of fragmentation functions with some reliability we would wish to operate $10^9$ consecutive CNOT gates. With the current 75 nanoseconds per gate 
this would put a basic fragmentation computation in the ballpark of 1 minute per shot (individual program run). The decoherence time is currently at the level of milliseconds at best, so factors $10^4$-$10^5$ have to be gained in maintaining quantum coherence. 
These projections will be discussed below in sections~\ref{sec:eos} and~\ref{sec:fragmentation} respectively.

Because implementing these many degrees of freedom is {\it very} expensive, our and other's calculations are very simple, limited to a few particles, and few momenta. The focus has been on testing and developing algorithms for a future quantum computer, not on the classical simulations per se.

\section{Lattice-Gauge Theory on a Quantum Computer}\label{sec:lattice}
Soon after Quantum Chromodynamics was established as the fundamental theory
underlying hadron physics, Lattice Gauge Theory (LGT), traced back~\cite{Wilson:1974sk} to Ken Wilson, emerged as a robust method with systematically (albeit slowly) improving uncertainties. It is natural, given their broad applicability beyond the original studies of confinement in the strong coupling regime, to attempt to employ quantum computing to fill the gaps where the application of LGT has ran into difficulty to arrive at phenomenologically successful applications. LGT is normally formulated in the Lagrangian, path--integral formalism, but such Euclidean computations are unnatural in Quantum Computing where time evolution is in real Minkowski space. Thus, the alternative Hamiltonian formulation of LGT is used.

LGT is a discretised representation of Quantum Chromodynamics on a finite spacetime grid of $n$ points with spacing $a$. 
In comparing with other formulations of QCD, LGT's formalism stands out because it is most often constrained to explicitly satisfy gauge invariance.
For this purpose, instead of using the gauge--dependent $A^a_\mu(y)$ fields whose quanta are the gluon partons, a change of variables~\cite{Rothe:1992nt,Gattringer:2010zz,Gross:2022hyw} to a series of \textit{link} variables
$U_{n\mu}$ named \textit{Wilson lines} is performed
\begin{equation}
    U_{n\mu} = \mathcal{P}\exp\left\{ig \int^{a\left(n+\hat{\mu}\right)}_{an} dy_{\nu} A_{\nu}(y)\right\}
   =\exp\left\{iga\sum_\alpha T^\alpha A^\alpha_{n\mu}\right\},
    \label{eq:Wline}
\end{equation}
where $a$ is the lattice spacing, $g$ the coupling constant and $\lambda^a = \frac{1}{2}T^\alpha$ the Gell-Mann matrices. The electric field operators $E^\alpha_{n\mu}$ are the conjugate variables of the $A^\alpha_{n\mu}$, defined as the integral of the gauge fields $A_\nu(y)$ over a path connecting $an$ and $a(n+\hat{\mu})$.  The Wilson line operators satisfy unitarity,
\begin{equation}
    U_{n\mu} = U\left(n,n + \hat{\mu}\right) = U\left(n + \hat{\mu},n\right)^{\dagger} = U\left(n + \hat{\mu},n\right)^{-1},
\end{equation}
and under a gauge transformation $\Omega$ on sites $n$ and $n+\hat{\mu}$ they change as 
\begin{equation}
    U'_{n\mu} = \Omega_n U_{n\mu}\Omega_{n+\hat{\mu}}^{-1}\ .
\end{equation}
This transformation can be exploited to obtain a manifestly gauge-invariant \textit{Wilson's plaquette} therefrom
\begin{equation}
        U_{n\ \mu\nu} =  U_{\square}  = \text{Tr}\left(U_{n\mu} U_{\left(n+\hat{\mu}\right)\nu} U_{\left(n+\hat{\nu}\right)\mu}^{\dagger} U_{n\nu}^{\dagger}\right)
\end{equation}
thanks to the cancellation of multiplied $\Omega \Omega^{-1}\dots$ factors upon expanding the product. The gauge dependence of the $A$ fields has been traded by a path (and grid) dependence of the plaquette $U_{n\ \mu\nu}$. The hope is that upon taking the continuum and large-volume limits one can reduce this dependence to be arbitrarily small.

The simplest discretisation of the QCD action is
the Wilson action
\begin{equation}
    S_{W} = \frac{3}{g^2} \sum_{n}\sum_{\mu\neq \nu}\left(1-\frac{1}{3}\text{Re}\left\{U_{n\mu\nu}\right\}\right),
\end{equation}
which can be seen to reproduce the correct continuum limit (the factor $3$ comes from the colour $SU(3)$ group).  
Fermion fields $\psi(x)$ become fields over the discretised set of nodes $\psi(x_n)$ or for short $\psi(n)$.
Using symmetric formulae for the derivatives
$\partial_{\mu}\psi_{\alpha}(n) = \frac{1}{2}\left[\psi_{\alpha}(n+\hat{\mu}) - \psi_{\alpha}(n-\hat{\mu})\right]$ 
 allows to discretise the Dirac action.
 
 An unavoidable difficulty is the appearance of fermion \textit{doublers} from contributions to the propagators coming from the edges of the Brillouin zones in momentum space, causing problems in implementing chiral symmetry. There are several ways of dealing with these doublers, e.g. \textit{Wilson}, \textit{twisted mass}, \textit{staggered} fermions, etc. We here focus on the last type, the ones most commonly used in today's formulations of LGT on quantum computers. The idea behind staggered fermions is to ``double'' the effective lattice spacing by distributing the fermions degrees of freedom so that the Brillouin zone is halved. This requires an extra $2^d$ points for each original lattice point, where $d$ is the space-time dimension under consideration\cite{Susskind:1976jm}. 
 
In toy models being studied in preparation for QCD, $d$ is fixed to $1+1$ (one timelike and one spacelike dimensions), and therefore there is an additional lattice site for each fermion mode, which can be used to encode quarks and antiquarks \cite{Farrell:2022wyt}. In 4 ($1+3$) dimensions,  there are four spinor components and $2^4=16$ sites per fermion mode, which give rise to non-physical \textit{tastes} (not to be confused with \textit{flavours})\cite{Gross:2022hyw}.

Space-time points (in lattice units, that is, with unit-cell's side-length of 1) can be labelled as $x_\mu = 2 n_\mu$ and the lattice points as $r_\mu = 2 n_\mu + \rho_\mu$, with $\rho$ a vector whose components are either $0$ or $1$.  New fields $\chi_\rho(x) = \chi(2n + \rho)$ can be introduced, related to the quark spinors by
\begin{equation}
    \psi^{l}_{\alpha}(x) = \mathcal{N}\sum_{\rho} \left(\Theta_\rho\right)_{\alpha l}\chi_{\rho}(x),
\end{equation}
where $\alpha$ and $l$ stand, respectively, for spinor and  ``taste'' indices. The $\left(\Theta_\rho\right)_{\alpha l}$ matrices are constructed from Dirac's $\gamma$, and $\mathcal{N}$ is a constant to ensure the correct continuum limit. The following \textit{staggered--fermion} action is then natural,
\begin{align}
    S_F = & \frac{1}{2} \sum_{n,\rho,\mu}\eta_\mu(\rho)\bar{\chi}(2n+\rho)\left[\chi(2n+\rho+\hat{\mu})-\chi(2n+\rho-\hat{\mu})\right] \nonumber\\
    +&M\sum_{n,\rho}\bar{\chi}(2n+\rho)\chi\left(2n+\rho\right),
\end{align}
where $\eta_\mu(\rho) = \eta_\mu(2n+\rho)$ is equal to $1$ or $-1$ depending on the binary values stored in $\rho$, i.e., on the lattice site within each space-time point $x$.
Finally, each field $\chi_{\rho}(n)$ can be promoted with a colour index 
so under a gauge transformation $\chi_{\rho}(n)\rightarrow \Omega_n \chi_{\rho}(n)$, rendering gauge invariant the following lattice action,
\begin{align}
    S_{QCD} = & S_W + \frac{1}{2} \sum_{n,\rho,\mu}\eta_\mu(\rho)\bar{\chi}(r)\left[U_{r\mu}\chi(r+\hat{\mu})-U^{\dagger}_{(r-\hat{\mu})\mu}\chi(r-\hat{\mu})\right] \nonumber\\
    + & M\sum_{n,\rho}\bar{\chi}(r)\chi\left(r\right)\ . \label{LQCD}
\end{align}

The standard LGT setup now proceeds by a Monte Carlo sampling of the path integral for QCD's Euclidean correlation functions employing the exponential of this action as its weight. But for quantum computer applications, the real--time Hamiltonian formalism is deployed, and to it we now turn.

\subsection{Hamiltonian formulations}
\label{subsec:KS}
Kogut \& Sussking (KS) Hamiltonian formulation of non-Abelian LGTs (HLGT) is based on an elegant rigid rotator analogy \cite{Kogut:1974ag}. The Weyl-gauge ($A^0 = 0$) Hamiltonian corresponding to Eq.~(\ref{LQCD}) is \cite{Kogut:1974ag,Ligterink:2000} 
\begin{align}
    H_{KS} & = \frac{g^2}{2a}\sum_{r,k}\sum_{a}E_{k}^{a}(r)E_{k}^{a}(r) + \frac{6}{a g^2}\text{Tr}\left\{1-\frac{1}{6}\left(U_{\square}+U_{\square}^{-1}\right)\right\} \nonumber\\
    & + \frac{1}{a}\sum_{r}\psi^{\dagger}(r)\,\frac{\vec{\sigma}\cdot \vec{n}}{i}\,U_{r\mu}\,\psi(r+\hat{\mu}) + m_0\sum_r (-1)^r \psi^{\dagger}(r)\psi(r), \label{KSH}
\end{align}
in terms of two--component spinors. As usual in a Hamiltonian approach, explicit covariance is lost. Additionally, in this underconstrained $A_0=0$ gauge, the remaining redundant degrees of freedom of gauge fields have to be expunged from the lattice: the KS Hamiltonian has an overcomplete {\it kinematic Hilbert space} and it has to be complemented with the $\textit{Gauss operator}$ to define physical states, see section~\ref{subsec:Weyl} below.

Key towards the simulation of Eq.~(\ref{KSH}) is the definition of a mapping between the Hilbert space of Chromodynamics and that of the quantum--computer as described in the next subsection.

\subsection{Mapping Field Theory to Quantum Computer Variables} \label{sec:encodingI}
\subsubsection{Fermion mappings}
Once the fermion degrees of freedom are properly distributed on the lattice, they can be mapped to a memory of qubits with fermionic--state occupation number 0 or 1, and therefore assignable to a single qubit. Spin and position degrees of freedom have already been taken into account by staggering. Then, encoding for example quark modes with two flavours and three colours would require $2\times3 = 6$ qubits per lattice site (for concrete examples~\cite{Farrell:2022wyt}  and further general discussion~\cite{Zohar:2014qma} about the representation of fermions in arbitrary HLGTs we refer to the literature). 

Creation and annihilation operators of these modes are then written in terms of lowering $\sigma^{-}$ and rising $\sigma^{+}$ operators, in turn decomposed in $\textit{Pauli strings}$ --tensor products of Pauli operators $X\equiv\sigma_x,Y\equiv\sigma_y,Z\equiv\sigma_z$ and the identity $\mathcal{I}$. Such a mapping is usually based on the Jordan-Wigner transformation\cite{Jordan:1928wi,Ortiz:2000gc}:
\begin{equation}
    b^{(\dagger)}_{j} = Z^{\otimes n-j-1}\otimes (\sigma^{+(-)})^j\otimes\mathcal{I}^{\otimes j-1},
\end{equation}
where the index $j$ runs over fermion modes. 

Alternative encodings are also possible~\cite{Bravyi:2000vfj}; in particular one can assign the modes of each quark (including position or momenta and spin) to a distinct qubit register~\cite{Abrams:1997gk,Kirby:2021ajp,Galvez-Viruet:2024hry}. We postpone a more detailed discussion to next sections since these encodings are not usually applied to HLGT fermions.

\subsubsection{Gauge-boson mappings}
The computer-discretised linear space corresponding to the gauge fields in Eq.~(\ref{KSH}) arises from direct products of single-link Hilbert spaces. We review here a construction in the electric-field basis due to Byrnes and Yamamoto\cite{Byrnes:2005qx}, although several other basis are possible~\cite{Zohar:2014qma}. 

The electric field operators $E^\alpha_{n\mu}$ (conjugate to the $A^\alpha_{n\mu}$ in Eq.~(\ref{eq:Wline})) and the Wilson lines satisfy the following equal time commutation relations
\begin{equation}
    \left[E^\alpha, U_{ij}\right] = T^\alpha_{ik} U_{kj},\,\,\left[E^\alpha,E^\beta\right] = if^{\alpha\beta\gamma}E^\gamma,
\end{equation}
where $f^{\alpha\beta\gamma}$ are the structure constants of the group: $\left[T^\alpha,T^\beta\right]= \frac{1}{2}i f^{\alpha\beta\gamma}T^\gamma$. Thus the electric fields $E$  are analogous to angular-momentum like variables and the Wilson lines to ladder operators. The Fock space of links can therefore be constructed by successive applications of Wilson lines to an initial state defined by $E^{\alpha}_{r,\mu}\left|0\right\rangle = 0$ on all links.

Link states are characterised by two sets of ``angular'' quantum numbers, one for each of the link's ends. For $SU(3)$ we have
\begin{equation}
    E^2\left|p,q,\lambda_L,\lambda_R\right\rangle = \frac{1}{3}\left(p^2+q^2+pq+3(p+q)\right)\left|p,q,\lambda_L,\lambda_R\right\rangle,
\end{equation}
where $p,q$ label the irreducible representation of $SU(3)$ and $\lambda_{L/R}$ is a shorthand for the quantum numbers that label the left and right ends of the link respectively. These quantum numbers can be defined in different ways; we here employ the representation 
of the colour group
mapped  from the flavour ($u$, $d$, $s$) triplet, adopting  the names ``isospin'' $T$ and ``hypercharge'' $Y$. The explicit state of the link is therefore
\begin{equation}
    \left|p,q,\lambda_L,\lambda_R\right\rangle = \left|p,q;T_L,T_{Lz},Y_{L};T_R,T_{Rz},Y_{R}\right\rangle.
    \label{eq:linkstate}
\end{equation}
The Wilson line operators are characterised by the link  quantum numbers and they are either in the fundamental ($p=1,q=0$) or adjoint ($p=0,q=1$) representations. Single-link states are therefore:
\begin{equation}
    \left|p,q,\lambda_L,\lambda_R;n+\hat{\mu}\right\rangle =\sqrt{\text{dim}(p,q)}\, U^{(p,q)}_{\lambda_L,\lambda_R;n+\hat{\mu}}\left|0;n+\hat{\mu}\right\rangle \ .
\end{equation}
By successive application of these Wilson-line operators all other link states can be generated; if the Wilson line has spin and hypercharge quantum number $(T_i,T^Z_i,Y_i)$ ($i=L,R$); and the link was initially in the $(p,q)$ representation with spin and hypercharge $(t_i,t_i^z,y_i)$, then the spin and hypercharge of the starting link and of the Wilson-line operator are added together 
\begin{equation}
    Y'_i = Y_i+y_i\ ;\\ \ \ \ \ \ \ \ \  T_i^{z'} = T_i^z+t_i^z,
\end{equation}
and the representation changes to a combination of $\left\{(p,q+1),\right.$ $(p+1,q-1)$, $\left.(p-1,q)\right\}$ or $\left\{(p+1,q),(p-1,q+1),(p,q-1)\right\}$ if the Wilson line was in the fundamental or adjoint representations respectively. The relative weights of each combination of quantum numbers depends on the Clebsh-Gordan coefficients of SU(3).

In order to build  an encoding for a quantum computer, each of the eight link quantum numbers of Eq.~(\ref{eq:linkstate}) must be encoded in a register of qubits. If the highest representation to be stored is the $(p_{max},q_{max})$ then two registers should have $\log_2(p_{max})$ and $\log_2(q_{max})$ qubits to store the different $p$ and $q$ values. The remaining quantum numbers must belong to the ranges
\begin{align}
    T_i& =\left\{0,\frac{1}{2},...,\frac{1}{2}(p+1)\right\} \\
    T_i^z&=\left\{-\frac{1}{2}(p+1),-\frac{1}{2}(p+1)+\frac{1}{2},...,\frac{1}{2}(p+1)\right\}\\
    Y_i&=\left\{-\frac{1}{3}(q+2p),-\frac{1}{3}(q+2p)+ \frac{1}{3},...,\frac{1}{3}(p+2q)\right\},
\end{align}
and each range's cardinality fixes the necessary number of qubits per quantum number. The Wilson line operators can then be implemented by appropriate combinations of lowering and raising operators on each of these quantum number registers\cite{Byrnes:2005qx}. 

The scaling of the needed number of qubits depends on the total number of terms in the Hamiltonian, which is proportional to the number of lattice sites $M$, to $d$, the space-time dimension, and to the number of qubits for each term $D$. This results in an overall scaling as $\mathcal{O}(MdD)$. 

Next, the scaling of the number of operations depends on the configuration of qubits in the actual device; if there were only nearest--neighbour interactions the scaling would be $\mathcal{O}(M^2d^2D)$, {\it i.e.}, quadratic in the lattice--site number, while with long-range interactions it can be reduced to be linear in the lattice sites $\mathcal{O}(Md)$.

Once the set of states and their corresponding mappings to a qubit memory are specified, a very non-trivial task remains. This is to find efficient decompositions of the unitary transformations between the encoded states in terms of some basic and universal set of quantum gates (usually single-qubit rotations and at least an entangling gate, such as the controlled-X, CNOT) and to $\textit{transpile}$ them into hardware-based operations.

\section{Particle-based encoding of Chromodynamics} \label{sec:particleregisters}

The encoding reviewed in this section~\cite{Galvez-Viruet:2024hry} is formulated from a few-body perspective in which the basic object is the particle. 
The quantum computer memory is organised in such a way that it resembles quantum field theory in the particle basis in second quantisation.
 
This encoding is competitive respect to, for example, the Jordan-Wigner encoding, when the number of particles is not too large but the number of momentum and other modes that each particle can have is large), it is appealing due to its intuitive use for particle-physics problems. In particular, it is constructed to formulate quantum field theory in light-front gauge, time-axial gauge, and Coulomb gauge.  
 
In contrast to particle encodings mentioned in the previous section~\ref{sec:encodingI} (among others), this encoding implements creation and annihilation operators that fulfil both commutation and anti-commutation relations up to boundary terms that emerge when the memory is completely occupied (and provided the memory is filled following a certain order). In particular, this implies that the Pauli principle among fermions is automatically implemented, but without restricting bosons.

A memory register is assigned to each particle active in a simulated process. Such register is a set of qubits that encode the quantum numbers of a particle $q$ and an additional qubit indicating the presence or absence of such particle (indicated by that qubit set to 1 or 0, respectively). If the presence/absence qubit, denoted by $P/A$ is set to zero in the register, the qubits are available. 

\subsection{Bosons}
\subsubsection{Single-particle boson operators} 

For a particle with momentum, spin and colour quantum numbers, the vacuum state is represented by the register:
\begin{eqnarray}
|\Omega\rangle 
\equiv
|0\rangle_{P/A} 
\otimes
|0\rangle_{\text{spin}} 
\otimes
|00\rangle_{\text{colour}}
\otimes
|0 \dots 0\rangle_{\text{momentum}}  
\end{eqnarray}
 
Encoding of spin and colour requires 1 and 2 qubits, respectively, whereas the $N_p$ values of 1-dimensional discretised momentum require $\log_2 N_p$ qubits [$(\log_2 N_p)^3$ in three dimensions].
 
Creation/annihilation operators $a^\dagger$/$a$ can be written in terms of \textit{set}/\textit{scrap}   and control  operators, respectively named  $\mathfrak{s}^\dagger$/$\mathfrak{s}$ and $\mathfrak{C}_{{a}{b}}$ ($a,b\in\{0,1\}$), as the product
\begin{eqnarray}
a^\dagger_{s,c,p}
\equiv
\mathfrak{C}_{10}
\otimes
\mathfrak{s}_s^\dagger
\otimes
\mathfrak{s}_c^\dagger
\otimes
\mathfrak{s}_p^\dagger\ .
\label{eq:adg}
\end{eqnarray}
The control operators therein act on the presence/absence qubit and fulfil the following relations
\begin{eqnarray}
\mathfrak{C}_{ij} |k\rangle
\es 
\delta_{jk}|i\rangle \ , \qquad
\mathfrak{C}_{ik}\mathfrak{C}_{lm}
\ = \ \delta_{kl} \mathfrak{C}_{im} \ .
\end{eqnarray}

When acting on the vacuum state $|\Omega\rangle$, Eq.~(\ref{eq:adg}) yields a one-particle state with spin $s$, colour $c$ and momentum $p$:
\begin{eqnarray}
a^\dagger_{s,c,p} |\Omega\rangle
&\equiv &
| 1 \rangle_{P/A}
\otimes
|s\rangle
\otimes
|c\rangle
\otimes
|p \rangle
\ = \ | 1scp\rangle\ .
\end{eqnarray}

The annihilation operators are then the Hermitian conjugates of the $a^\dagger$ operators
$a_{s,c,q} = (a_{s,c,q})^\dagger$.

\subsubsection{Two-particle boson operators}
\label{twopartile}

The vacuum state in a two-particle memory space is represented using two registers:
\begin{eqnarray}
|\Omega\rangle 
\ = \ |\Omega\rangle_2 \otimes |\Omega\rangle_1
\equiv
|0\rangle_{P/A} 
|0 \dots 0\rangle_{\text{momentum}}  
\otimes
|0\rangle_{P/A} 
|0 \dots 0\rangle_{\text{momentum}}\ ,
\end{eqnarray}
where spin and colour quantum numbers are now hidden  for simplicity of reading.

Particle operators acting on a two-register memory are represented with a ``(2)" superscript. 
For instance, a particle creation operator acting on the vacuum reads
\begin{eqnarray}
a_{q_1}^{(2)\dagger} |\Omega\rangle
= 
|q_1\rangle 
\equiv
|\Omega\rangle_2 \otimes |1 q_1\rangle_1\ .
\end{eqnarray}
In order to describe one-particle in a two-register memory, we resolve the ambiguity of choosing the active register by the convention that registers always activate from right to left. This excludes from the Fock space  states such as $|1 p\rangle_2 \otimes |\Omega\rangle_1$.

The action of the creation operators needs to yield states representing the abstract symmetry under boson exchange. So if a register on a two-register memory is occupied by one boson, and a creation operator acts, Bose statistics requires
\begin{eqnarray}
a_{q_1}^{(2)\dagger} |\Omega\rangle_2
\otimes |1p_1\rangle_1
\equiv
{1 \over \sqrt{2}} 
(|1 q_1\rangle_2 \otimes |1 p_1\rangle_1
+ 
|1 p_1\rangle_2 \otimes |1 q_1\rangle_1
)\ ;
\end{eqnarray}
but if instead  both registers already contain a boson, no additional particle can be encoded, and we adopt the convention that the creation operator send the two-particle state to null: 
\begin{eqnarray}
a_{q_1}^{(2)\dagger} |1p_2\rangle_2 |1p_1\rangle_1 = 0 \ .
\label{fullmemoryzero}
\end{eqnarray}
As a consequence of the limited computer memory (only two registers), the Fock space is truncated, introducing a discretisation error in the second quantised problem. With an increasing number of particle registers, this inconvenience is pushed up far from the few--body spectrum of the theory.

We split the particle operator over a two-register memory into two suboperators
\begin{eqnarray}
a_{q_1,1}^{(2)\dagger} 
\es 
(\mathfrak{C}_{00}\otimes \mathfrak{i})_2 \otimes (\mathfrak{C}_{10}\otimes \mathfrak{s}_{q_1}^\dagger)_1 \ , \\
a_{q_1,2}^{(2)\dagger} 
\es 
(\mathfrak{C}_{10}\otimes \mathfrak{s}_{q_1}^\dagger)_2\otimes (\mathfrak{C}_{11}\otimes \mathfrak{i})_1 \ ,
\label{2bodyauxiliary}
\end{eqnarray}
the first (second) operator creates a particle with quantum numbers $q_1$ in the first (second) register; operations over the second register leave the first register untouched acting with the identity $\mathfrak{i}$, assuming it to be occupied. 

The creation operator is then a sum of the first suboperator plus the second one with a symmetriser (since it produces a doubly-occupied memory), 
\begin{equation}
a_{q_{1}}^{(2)\dagger} = a^{(2)\dagger}_{q_{1},1}+ \frac{1}{\sqrt{2}}\left(\outeridentity\otimes \outeridentity+\mathcal{P}_{21}\right)a^{(2)\dagger}_{q_{1},2}\ .
\label{2bodycreator}
\end{equation}
Here, the \textit{permutation operator} $\mathcal{P}_{21}$ swaps   quantum numbers among the registers
and  $\outeridentity$ is the \textit{register-level identity} which leaves the quantum numbers in the corresponding register unchanged. In what follows we will name
\begin{eqnarray}
 \frac{1}{\sqrt{2}}\left(\outeridentity\otimes \outeridentity+\mathcal{P}_{21}\right) 
 & \equiv &
 \mathcal{S}_2\ ,
\end{eqnarray}
a 2-particle \textit{symmetriser}.
The operators then verfy the boson commutation relations
\begin{equation}
\left[a_{q_{1}}a^{\dagger}_{q_{2}}\right] = a_{q_{1}}a^{\dagger}_{q_{2}}-a^{\dagger}_{q_{2}}a_{q_{1}} = \delta_{q_{1}q_{2}}\ .
\end{equation}
when acting on the vacuum and one particle states, but they fail when the memory is full, as trying to create a third particle on the two-particle register is not possible by construction, so a creation operator acting on two-particle operators yields 0 as per Eq.~(\ref{fullmemoryzero}). 
Therefore, the correct commutation relations are fulfiled up to a boundary term that activates when the memory is totally filled, in this case by two bosons.

\subsubsection{$n$-particle operators}
The $n$-register vacuum is written as 
\begin{equation}
    \ket{\Omega}\equiv\ket{\Omega}_{n}\otimes ... \otimes \ket{\Omega}_{1} = \underbrace{\left(\ket{0}_{P/A}\ket{0...0}\right)_n\otimes ... \otimes\left(\ket{0}_{P/A}\ket{0...0}\right)_1}_{n}\ ,
    \label{def:nReg-vacuum}
\end{equation}
Generalising section~\ref{twopartile}, $n$-register operators are constructed by means of projectors (control), $\mathfrak{C}_{ij}=|i\rangle\langle j|$, identities $\mathfrak{i}$ and set/scrap operators $\mathfrak{s}^{\dagger}_p/\mathfrak{s}_p$. The operator that activates the first register is
\begin{equation}
a^{(n)\dagger}_{p,1} \equiv \left(\ketbrac{0}{0}\otimes\innerid\right)_n...\left(\ketbrac{0}{0}\otimes\innerid\right)_2\left(\ketbrac{1}{0}\otimes\mathfrak{s}^{\dagger}_{p}\right)_1\ ,
\end{equation}
indeed
\begin{eqnarray}
a^{(n)\dagger}_{p,1}\ket{\Omega}  
\es 
\left(\ket{0}_{P/A}\ket{0...0}\right)_n...\left(\ket{1}_{P/A}\ket{p}\right)_1\ .
\end{eqnarray}
 
The most efficient way to implement the symmetrisation for more than two registers is to proceed in a stepwise fashion, assuming  previously symmetrised $n-1$ registers and requiring the newly filled register to also be in a symmetric wavefunction with the remaining ones. Hence, the $n$-particle symmetriser\footnote{This symmetriser is not idempotent, but satisfies  $S^2_{n\leftarrow n-1} =
\sqrt{n}\ S_{n\leftarrow n-1}$ instead.}
\begin{equation}
    \mathcal{S}_{n\leftarrow n-1} \equiv \frac{1}{\sqrt{n}}\left(\outeridentity^{\otimes n}+\mathcal{P}_{n(n-1)}+...+\mathcal{P}_{n2}+\mathcal{P}_{n1}\right)\ .
\label{def:step-symmetriser}
\end{equation}

Operators $a_{q,j}^{(n)\dagger}$ that create a particle on the $j$th register are defined analogously to (cf.~\ref{2bodyauxiliary}), and the total creation operators with the required commutation relations are compactly written as

\begin{equation}
a^{(n)\dagger }_{q} = \sum^{n}_{i = 1}a^{(n)\dagger }_{q,i}  =  \sum^{n}_{i = 1}  \mathcal{S}_{i\leftarrow(i-1)} \cdot \projector{n-i}{0}\otimes \left(\ketbrac{1}{0}\otimes \mathfrak{s}^{\dagger}_{q}\right)_{i} \otimes \,\projector{i-1}{i-1}\ ,
    \label{def:nReg-bosoncreation}
\end{equation}
and
\begin{equation}
    a^{(n)\dagger }_{p} \left(\ket{1p_{n}}\ket{1p_{n-1}}...\ket{1p_{1}}\right)_{\mathcal{S}} = 0\ .
\end{equation}

Annihilation operators are the adjoint of creation operators, therefore:
\begin{equation}
a^{(n)}_{q} = \sum^{n}_{j=1}a^{(n)}_{q,j} = \sum^{n}_{j=1} \projector{n-j}{0}\otimes \left(\ketbrac{0}{1}\otimes \qscrap{p}\right)_{j} \otimes \,\projector{j-1}{j-1} \cdot\mathcal{S}_{j\leftarrow(j-1)}\ .
\label{def:nReg-bosonannihilator}
\end{equation}
The action of the operator over an $i$-particle symmetric state is
\begin{align}
a^{(n)}_{q} & \ket{\Omega}_n...\ket{\Omega}_{i+1}\left(\ket{1p_i}_i...\ket{1p_1}_1\right)_S  =  \sum^{n}_{j = 1}a^{(n)}_{q,j}\ket{\Omega}_n...\ket{\Omega}_{i+1}\left(\ket{1p_i}_i...\ket{1p_1}_1\right)_S\nonumber \\
& = \sum^{i}_{l=1}\delta_{p p_l}\ket{\Omega}^{\otimes n-i+1}\left(\ket{1p_i}_{i-1}...\ket{1p_{l+1}}_{l}\ket{1p_{l-1}}_{l-1}...\ket{1p_1}_1\right)_S\ ,
\end{align}
which would be impossible to satisfy with operators acting on single registers (non-trivially) if the memory was not explicitly symmetrised.

These operators satisfy the following commutation relations in the Hilbert subspace of symmetric states
\begin{align}
    \left[a^{(n)}_{\rho},a^{(n)\dagger}_{\eta}\right]  &= \delta_{\rho\eta}\left(\ketbrac{0}{0}\otimes\innerid\right)_{n}\otimes \mathbb{I}^{(n-1)}\ \nonumber \\&  -\ \,S_{n\leftarrow n-1}\cdot \left(\ketbrac{1}{1}\otimes \mathfrak{s}^{\dagger}_{\rho}\mathfrak{s}_{\eta}\right)_n \otimes\projector{n-1}{n-1}\cdot S_{n\leftarrow n-1} \ .
    \label{eq:nReg-commutation}
\end{align}
\paragraph{Number operator}.  These definitions can be applied to any operator written in terms of creators and annihilators, for the \textit{number operator} we have
\begin{align}
    \hat{N}^{(n)} & := \sum_{p} a^{(n)\dagger}_{p}a^{(n)}_{p} = \sum_{p}\sum_{j,j'} a^{(n)\dagger}_{p,j}a^{(n)}_{p,j'}\nonumber  \\
    & = \sum_{j} \mathcal{S}_{j\leftarrow j-1} \cdot \projector{n-j}{0}\otimes \left(\ketbrac{1}{1}\otimes \sum_{p} \qset{p}\qscrap{p}\right)_{j} \otimes \,\projector{j-1}{j-1} \cdot \mathcal{S}_{j\leftarrow j-1} \nonumber \\
    & = \sum_{j} \mathcal{S}_{j\leftarrow j-1} \cdot \projector{n}{j} \cdot \mathcal{S}_{j\leftarrow j-1}\nonumber \\
    &= \sum_{j} \,j\,\projector{n}{j}\ ,
    \label{Number-operator}
\end{align}
where in the last line we assumed a symmetric memory so that $ \mathcal{S}_{j\leftarrow j-1}$ (c.f.~\ref{def:step-symmetriser})
simplify to $\sqrt{j}$.
\subsection{Fermions}

Because the conceptual architecture is very similar, we only briefly touch on anticommuting operators. We change the recursive step-symmetriser operators to step-antisymmetrisers
\begin{equation}
    \stepasym{n} = \frac{1}{\sqrt{n}}\left(\outeridentity^{\otimes n}-\mathcal{P}_{n(n-1)}-\mathcal{P}_{n(n-2)}-... -\mathcal{P}_{n2}-\mathcal{P}_{n1}\right)\ ,
    \label{def:step-asym}
\end{equation}
Creation and annihilation operators of fermions are then constructed using analogous arguments, and can be written, respectively as
\begin{equation}
b^{(n)\dagger }_{q} = \sum^{n}_{j = 1}b^{(n)\dagger }_{q,j} =  \sum^{n}_{j = 1}\stepasym{j}\cdot \projector{n-j}{0}\otimes \underbrace{\left(\ketbrac{1}{0}\otimes \mathfrak{s}^{\dagger}_{q}\right)_{j}}
\otimes \,\projector{j-1}{j-1}\ ,
    \label{def:nReg-fermioncreation}
\end{equation}
and
\begin{equation}
b^{(n)}_{q} = \sum^{n}_{j=1}b^{(n)}_{q,j} = \sum^{n}_{j=1} \projector{n-j}{0}\otimes \left(\ketbrac{0}{1}\otimes \qscrap{p}\right)_{j}
\otimes \,\projector{j-1}{j-1} \cdot \mathcal{A}_{j\leftarrow j-1} \ .
\label{def:nReg-fermionannihilation}
\end{equation}

They now fulfil the required anticommutation relations
\begin{eqnarray}
\left\{b^{(n)}_{q_{1}},b^{(n)\dagger}_{q_{2}}\right\}\  
\es \delta_{q_{1},q_{2}}\left(\ketbrac{0}{0}\otimes\innerid\right)_{n}\otimes \sum^{n-1}_{j=0}\projector{n-1}{j}\ \np
\ \mathcal{A}_{n\leftarrow n-1}\cdot \left(\ketbrac{1}{1}\otimes \mathfrak{s}^{\dagger}_{q_{1}}\mathfrak{s}_{q_{2}}\right)_n\otimes\projector{n-1}{n-1}\cdot \mathcal{A}_{n\leftarrow n-1}\ ,
    \label{eq:nReg-anticommutation}
\end{eqnarray}    
again, up to a boundary term when the memory is full.

\subsection{Quantum evolution}
Time evolution is one important quantum phenomenon that we can represent on a quantum computer but not on an Euclidean lattice formulation. Exponentiation of the Hamiltonian needs to be carried out, entailing products of creation and annihilation operators.

\begin{figure}
    \centering
    \includegraphics[width=0.26\linewidth]{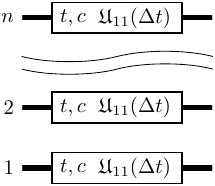}
    \includegraphics[width=0.6\linewidth]{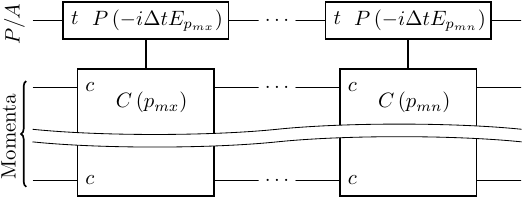}
    \caption{ Left: Circuit implementing the free-evolution term of Eq.~(\ref{eq:free-evolution}); registers are represented by thick lines.  Right: Implementation of $\mathfrak{U}_{11}$ in terms of controlled phase gates: $C(p)$. On a gate, t denotes target registers, whereas $c/ac$ denotes control/anticontrol registers. Copied from~\cite{Galvez-Viruet:2024hry} under the terms of the Creative Commons License 4.0 (http://creativecommons.org/licenses/by/4.0/)}
    \label{fig:U11schema}
\end{figure}

A particle-number-preserving unitary operator can be compactly written  as
\begin{equation}
\mathcal{U}^{f}_{11}\left(\Delta t\right) = \projector{n}{0} + \sum^{n}_{i=1}\projector{n-i}{0}\prod^{1}_{k=i}\otimes\left(\ketbrac{1}{1}\otimes  \mathfrak{U}_{11}\left(\Delta t\right)\right)_{k}\ ,
\label{eq:free-evolution}
\end{equation}
where $\mathfrak{U}\left(\Delta t\right)$ is an auxiliary register-level exponentiation (it can be seen as a composite gauge) of the form
\begin{equation}
\mathfrak{U}_{11}(\Delta t)\equiv \exp\left[-i\Delta t \sum_{q}E_q \qset{q}\qscrap{q}\right]\ .
\label{def:free-evolution-onregister}
\end{equation}
A naïve scaling following the implementation of Fig.~\ref{fig:U11schema} gives $\mathcal{O}(n\,N_p\,\log N_p)$, since there is a control over each of the $N_p$ possible configurations\cite{Galvez-Viruet:2024hry}. An implementation based on combination of Pauli strings can drastically reduce this cost, but the detailed scaling depends then on the values of the coefficient (in this case $E_p$) which are different over the different configurations.

For interaction terms that change the values of momenta we need to define a discretisation of $p$; for instance, the equidistant grid
\begin{equation}
    \left\{p_{min},...,p_{-1},p_0 = 0,p_1,...,p_{max}\right\}\ ,
    \label{Momentum-discretisation}
\end{equation}
with  $p_{l} = p_0+l\Delta = l\Delta$, $l\in \left\{\Lambda_{min},...,\Lambda_{max}\right\}$ and $|\left\{\Lambda_{min},...,\Lambda_{max}\right\}|=N_p$, for simplicity we keep $-\Lambda_{min} = \Lambda_{max} = \Lambda $. Negative values of $p$ are tagged in the storage by assigning a negative sign to the integer subindex.

The particle-number-conserving two-to-two momentum exchanges exponentiates to the unitary operator
\begin{equation} \label{4bodypotential}
\mathcal{U}^{f}_{22}(\Delta t) = \exp \left[-i \Delta t\sum^{\Lambda}_{\xi=-\Lambda}\left(\lambda_{\xi} \sum_{q,p} b_{q+\xi\Delta}^{\dagger} b^{\dagger}_{p} b_{p+\xi \Delta} b_{q}+h.c.\right)\right] \equiv \exp \left(-i\Delta t h^{f}_{22}\right)\ ,
\end{equation} 
where $\lambda_\xi$ represents the coefficients in front of the creation and annihilation operators in the Hamiltonian  and controls the probability or intensity of the exchange interaction. The limits of sums should be chosen so that $s=q+\xi \Delta = q + p_\xi$  and $r=p+\xi\Delta=p+p_\xi$.


When considering Hamiltonian interaction terms that alter the number of particles, we require a dynamical memory, which changes the number of active registers.
Similarly, we can define tadpoles, emission or absorption of particles, etc. For instance, the following operator stands for the evolution associated to the Hamiltonian term corresponding to the quark-gluon vertex
\begin{equation}
     \mathcal{U}_{21}(\Delta t, \lambda) = \exp\left[-i\Delta t \sum_{r,m}\lambda_{\xi}\left(a^{\dagger}_{p_\xi}b^{\dagger}_{p_r} b_{p_m}+h.c.\right)\right]_{\xi=m-r}\ ,
     \label{eq:splitting-term}
\end{equation}
where $p_\xi$, $p_r$, and $p_m$ are momenta encoded in the grid defined by Eq.~(\ref{Momentum-discretisation}) and following text. The exponentiation of such terms require a few bookkeeping steps that can be found in\cite{Galvez-Viruet:2024hry}.

\newpage
\section{Gauge fixing}

The gauge redundancy in Yang-Mills theory or Chromodynamics forces one to select a gauge when fixing the physical states in any Hamiltonian formulation. First and foremost we discuss equal--time quantisation in subsection~\ref{subsec:Weyl}. Then we turn to light--front quantisation in subsection~\ref{subsec:lightfront}.

With a bird's eye view, unless one fixes Coulomb gauge which is a complete gauge fixing with an entirely physical Fock space but a convoluted Hamiltonian necessitating approximation~\cite{Szczepaniak:2001rg,Reinhardt:2017pyr,Cotanch:2010bq,Llanes-Estrada:2000cdq}, the computer will often be approximating a redundant Fock space on which Gauss's law needs to be employed to project over a physical instance. The latter evolution with the gauge-invariant Hamiltonian should in principle remain within that projection, but truncation and creeping random errors can produce wavefunction components which are redundant gauge copies or worse; a proposal which we find interesting to ameliorate this gauge drift has been put forward by Ball and Cohen~\cite{Ball:2024xmw}, and it is to repeatedly project the gauge at short time intervals, and exploit the quantum Zeno effect that blocks the evolution of the measured observable.

\subsection{Time-axial gauge}\label{subsec:Weyl}
Coulomb gauge~\cite{Christ:1980ku,Cucchieri:2000gu} makes implementing the Hilbert space quite straightforward: there are only two transverse gluon creation operators  and their action on any physical state, for example the vacuum, produces another physical state. In exchange, the Hamiltonian is convoluted, as untangling the gauge redundancy brings in a kernel containing the inverse of an interacting differential operator which depends on the field operators $A^a$ themselves, complicating the simpler Coulomb potential of Electrodynamics.  

The reasonable alternative option which we (and many others) pursue in the context of quantum computing is to, instead, fix the time--axial gauge condition ${\bf A}^0=0$. This Weyl gauge presents a simple Hamiltonian, but in exchange the gauge redundancy has not been completely eliminated (the time--axial choice is one condition, when however two of the components of the gauge--field four vector need to be resolved in terms of the two physical ones).
Therefore, the (multi) gluon states that the three spatial $\vec{A}$ components generate need to be trimmed off unphysical ones. This is achieved by imposing Gauss's law as a constraint,
\begin{equation}
    0 = \mathcal{G}^a(r)\left|\psi\right\rangle_\text{phys} = \left(\vec{\nabla}\cdot\vec{E}^a(r) - \rho_G(r) - \rho_F(r)\right)\left|\psi\right\rangle_\text{phys}\ .
\end{equation}
Here, $\rho_F = \psi^{\dagger}T^a\psi$ is the fermion colour density (taken at each lattice site in the case of the  Kogut-S\"usskind formulation above). In turn, $\rho_G$ is its non-Abelian counterpart due to the gluon colour charge, and in the lattice formulation,  $\rho_G= \frac{1}{2}\sum_{\mu}Q_{r\mu}$ is the charge carried by the link that starts at $r$ and ends at $r+\hat{\mu}$. 

It is not obvious {\it a priori} that any of the states in the discretised ``kinematic'' Hilbert space actually satisfies Gauss's condition exactly. Instead, a common procedure is to minimise $\langle \mathcal{G}^2 \rangle$
over this kinematic space and expect that, in the continuum limit, the minimisation leads to the exact satisfaction of Gauss's law. 
There are here pitfalls which need to be avoided. For example, if the Fock space/ Hilbert space is defined in terms of normalizable wavefunctions only, as pertains to bound state problems, for example, the 
limiting procedure could lead to those states satisfying Gauss's law to fall outside that space. Gauss's law would then have to be applied, in a weaker sense, to its algebraic dual~\cite{Thiemann:2006cf} set (loosely, Dirac bracs) $\langle \psi'|$ so that the matrix elements of any observables $\langle \psi' | \mathcal{O} \psi\rangle $ would in effect comply with the gauge restriction.

\subsection{ Light-front gauge} \label{subsec:lightfront}


The light-front gauge is defined by $A^{a+}=0$. With this condition, the classical equations of motion lead to $A^{a-}$ being fixed by a constraint 
\begin{eqnarray}
\partial^+ A^{a-} 
\es
 2D^{ab\perp} A^{b\perp} 
-\frac{2}{\partial^+} j^{a+}\ ,
\label{Aminus}
\end{eqnarray}
and thus not being quantised, there remaining the two physical degrees of freedom $A^{a\perp}$.

The free term $\partial$ in the covariant derivative $D$ of Eq.~(\ref{Aminus}) is independent of the coupling constant $g$; looking at its purely Yang-Mills part (without the quark colour current $j$) it is customary to define $\tilde A^-$ as
\begin{eqnarray}
\tilde A^- 
\es
{1\over \partial^+} 2\partial^\perp A^\perp\ ,
\label{AminusTilde}
\end{eqnarray}
and include its second piece, 
$- {2 \over \partial^{+2}} i g[\partial^+ , A^\perp]$ in the interaction part of the Hamiltonian. In this document, this convention is taken, but the tildes are removed to simplify the notation.

The light-front gauge is well suited for front-form formulations of Hamiltonian dynamics~\cite{Dirac:1949cp} (see~\cite{Brodsky:1997de} for an extended review), where the quantisation surface in quantum theory--or the specifications of initial conditions in classical theory--is defined by $x^+:=t + z=0$, where $x^+$ is referred to as the light-front time.
The quantum dynamical evolution then proceeds along $x^+$:
\begin{eqnarray}
    \exp[- i \hat P^- x^+] \ ,
\end{eqnarray}
with the \textit{front-form Hamiltonian} $\hat P^-$ defined as the conserved Noether charge under front-form space-time translations. 

Light-front dynamics is useful for several reasons. First, conservation of momentum at interaction vertices, together with the positivity of the front-form momentum $p^+=p_1^+ + p_3^+$ precludes interactions with only creation or only annihilation operators, thereby simplifying the structure of the Hamiltonian\footnote{Of course this condition does not apply to only-gluon interactions, since in this case $p^+$ can vanish. This exception can, however, be circumvented by employing a regularisation procedure that assumes gluons to have an infinitesimally small mass~\cite{Glazek:2018kvx,Galvez-Viruet:2023bid}.}. Relativity requires preserving the commutation relations of the 10 generators of the Poincar\'e group. In a front form representation 7 of these generators are interaction free, one of them being the boost along $z$ direction, which in this form has a particularly simple structure. The front-form is therefore specially convenient for describing processes that involve a preferred direction of motion, as is the case in deep inelastic scattering. 

Two main approaches are used to perform numerical calculation in Light-Front QCD, which can be simulated on a quantum computer: Discretised Light-Cone Quantisation (DLCQ)~\cite{Pauli:1985ps,Pauli:1985pv} and Basis Light-Front Quantisation (BLFQ)~\cite{Vary:2009gt}, the latter originated from the former. 
DLCQ discretises the fundamental fields on the space-time, providing a suitable framework for simulating \textit{ab initio} fundamental interactions, whereas BLFQ adopts a basis function representation and is well suited for formulating effective theories and computing static hadron observables~\cite{Kreshchuk:2020dla}. 

Light-front wave functions have been obtained and used to calculate several pion observables~\cite{Kreshchuk:2020kcz,Kreshchuk:2020aiq} on a quantum computer by means of BLFQ. Parton distribution functions and form factors~\cite{Kreshchuk:2020dla,Kreshchuk:2020aiq} have been simulated using DLCQ, while the first efficient quantum simulations of QCD jets were obtained within an approach based on both of them~\cite{Barata:2023clv,Qian:2024gph}.

\section{Progress towards the Equation of State of Neutron Stars} \label{sec:eos}
One of the most anticipated applications of quantum computers is the characterisation of many-body systems from the simulation (whether digital or analogue) of their basic constituents and interactions. Current research mainly focuses  on the calculation of the partition function of canonical ensembles, which depends on temperature $T$, number of particles $N$
and volume $V$ (omitted):
\begin{equation}
    Z(\beta) = \text{Tr}\,e^{-\beta H}\ ,
    \label{Canonical}
\end{equation}
where $\beta = 1/(k_B T)$ and $H$  is the Hamiltonian of the system.

A common method considers the extension of the Hamiltonian parameters $h$ to the complex plane and the characterisation of the partition function by its complex zeros ({\it e.g.} those of Lee and Yang for the grand-canonical partition function~\cite{Yang:1952be,Lee:1952ig} or of Fisher\cite{Fischer:1965rna} for the canonical one as function of the inverse temperature $\beta$). The behavior of these zeros can also help elucidate the existence of phase transitions: they reside in the complex $\beta$ plane because $Z$ is a sum of positive terms, but as they approach the real axis they can cause nonanalyticities of thermodynamic functions. 

We are not yet aware of any computation approximating Chromodynamics; however, 
this procedure has been proven on an ion-trap quantum computer for the XXY model with anisotropic couplings\cite{Francis:2021}: 
\begin{equation}
    H_s = J\left(\sum_{i}\sigma^x_i\sigma^x_{i+1}+\sigma^y_i\sigma^y_{i+1}\right) + J_z\sum_{i}\sigma^z_i\sigma^z_{i+1}\ ,
\end{equation}
which exhibits a phase transition between an $XY$ phase ($|J|>|J_z|$) and an Ising phase ($|J|<|J_z|$). The Hamiltonian is combined with an external magnetic field term with complex coefficient
\begin{equation}
    H_B = (h_r+ih_i)\sum_i\sigma^z_i\ ,
\end{equation}
so that the complete partition function is 
\begin{equation}
    Z(\beta) = \text{Tr}\exp\left(-\beta H_0-i\beta h_i\sum_j\sigma^z_j\right)\ ,
\end{equation}
where $H_0 = H_s+\text{Re}(H_B)$. The zeros of the partition function can then be obtained by first preparing in the quantum computer a thermal state  with distribution $P(\beta) = e^{-\beta H_0}/Z(\beta,h_i=0)$ and then adding the ``imaginary part'' by coupling and auxiliary qubit in the state $\left|+\right\rangle=\frac{1}{\sqrt{2}}\left(\left|0\right\rangle+\left|1\right\rangle\right)$ and applying the unitary gate $U=\exp(-i\beta h_i\,\sigma^z_{\text{aux}}\otimes\sum_i\sigma^z_i)$. After evolution, the complex partition function can then be obtained by measuring the auxiliary qubit. The zeros obtained in this way are related to the phase change. The procedure was tested with an ion-trap quantum computer with two sites, for which two zeros are expected; in the XY-phase they appear for $h_r=0$ and $h_i$ symmetric about $\pi/(2\beta)$, while in the Ising phase the behavior is reversed, the zeros correspond to $h_r$ symmetric about 0 and $h_i=0$. The quantum computer can accurately predict this behavior. 

The most resource-intensive part of the algorithm is the preparation of the thermal state, whose realisation on quantum computers is a field of investigation all by itself. The authors of this extant work~\cite{Francis:2021} use the thermofield double (TFD) method for quantum computers\cite{Zhu:2019bri}; whose first step is to variationally prepare the state
\begin{equation}
    \left|\text{TFD}\,(\beta)\right\rangle = \frac{1}{\sqrt{Z(\beta)}}\sum_n e^{-\beta E_n/2}\left|n\right\rangle_A\left|n'\right\rangle_B\ ,
\end{equation}
in an extended Hilbert space with subsystems $A$ and $B$ with $\left|n\right\rangle$ and $E_n$ the eigenstates and eigenenergies of $A$ and $\left|n'\right\rangle = \otimes_jY_j\left|n\right\rangle$. Subsystem $B$ is then traced out, so that at the end subsystem $A$ is left in the thermal state with density matrix $\rho_A = e^{-\beta H_A}/Z(\beta)$. This procedure has recently been used \cite{Than:2025} to prepare thermal states of the Kogut-Sussking Hamiltonian (1+1D QCD, see sec. \ref{subsec:KS}) on a chain of trapped-ion qubits. The translational degrees of freedom in the y-direction are entangled with the ion's internal degrees of freedom storing the qubits in a way that depends on a parameter $\theta_i$ for the  $i$th ion:
\begin{equation}
    \left|\psi(\theta_i)\right\rangle = \cos\left(\theta_i/2\right)\left|0,0\right\rangle + \sin\left(\theta_i/2\right)\left|1,1\right\rangle\ ,
\end{equation}
where the computational basis is denoted by $\left|\text{spin},\text{motion}\right\rangle$. Tracing out the translational degrees of freedom, the memory is left in a mixture of bit strings $\left|j\right\rangle\equiv\left|j_1,j_2...j_N\right\rangle$ with probabilities $p_j(\vec{\theta}) = \prod^{N}_{i=1}p_{j_i}(\theta_i)$. This procedure is then used as part of the VQE algorithm to measure the chiral condensate $\hat{\chi}$ over states that minimize the free energy $F = E-TS$. Colour neutrality is then enforced using projection operator $\hat{K}$ to the singlet subspace. The expectation values $\left\langle\hat{O}\right\rangle_O$ over singlet subspaces are then defined as 
\begin{equation}
    \left\langle\right.\hat{O}\left.\right\rangle_O = \frac{\left\langle\right.\hat{O}\hat{K}\left.\right\rangle }{\left\langle\right.\hat{K}\left.\right\rangle}\ .
\end{equation}

Another popular method for thermal state preparation is the quantum imaginary-time evolution algorithm (QITE)\cite{Motta:2020} based on the fact that, if a state $\left|\Phi(\beta)\right\rangle$ satisfies
the imaginary-time Schrödinger equation
\begin{equation}
    -\partial_\beta\left|\Phi(\beta)\right\rangle = \hat{H}\left|\Phi(\beta)\right\rangle\ ,
\end{equation}
then 
\begin{equation}
    \left|\Psi_0\right\rangle = \lim_{\beta\rightarrow \infty}\frac{\left|\Phi(\beta)\right\rangle}{||\left|\Phi(\beta)\right\rangle||}\ ,
\end{equation}
converges (at $T=0$) to the ground-state of the Hamiltonian, always  provided $\left\langle\Phi(0)\,\right|\left.\Psi_0\right\rangle\neq 0$. The implementation on a quantum computer then rests on the ability to efficiently unitarise the non-unitary transformations $e^{-\beta H}$. It has been shown\cite{Motta:2020} that this can be done provided the Hamiltonian is geometrically k-local, which means that the Hamiltonian can be written as a sum with terms each acting on k-neighbouring qubits $H=\sum_m h[m]$. 
This is expected to happen in Quantum Chromodynamics since the Hamiltonian is a polynomial of degree 4 in the fields, however, some gauges can in principle induce long-range correlations. This needs further case-by-case study.

A Trotter decomposition can then be used to evolve with each of the $h[m]$ over a small ``time'' $\Delta\tau$ with $\beta=N\Delta\tau$,  $\exp\left\{-\beta\, h[m]\right\}$ is then realised approximately by repeatedly acting with  $\exp\left\{-i\Delta\tau A[m]\right\}$, with $A$ a Hermitian matrix decomposed in terms of Pauli strings
\begin{equation}
    A[m] = \sum_m a[m]_{i_1,...,i_k}\sigma_{i_1}...\sigma_{i_k} = \sum_m a[m]_{I}\sigma_{I}\ ,
\end{equation}
acting on the $k$-neighbouring qubits. The coefficients $a[m]$ are then found by solving the linear system $S_{I,I'}\,a[m] = b_{I}$ with 
\begin{equation}
    S_{I,I'} = \left\langle \Psi\right|\sigma^{\dagger}_I\sigma_{I'}\left|\Psi\right\rangle\ ,\,\,\,\,b_I = \frac{-i}{\sqrt{c}}\left\langle \Psi\right|\sigma^{\dagger}_Ih[m]\left|\Psi\right\rangle\ ,
\end{equation}
with $c = 1-2\Delta\tau\left\langle \Psi\right|h[m]\left|\Psi\right\rangle + \mathcal{O}(\Delta\tau^2)$. Finally, as the procedure is carried over and correlations are built over the system, the support of each of the $A[m]$ operators have to be extended to cover the new correlation lengths, so the efficiency of this protocols lies ultimately on the finiteness and smallness of the correlation length C: for geometrically k-local Hamiltonians in $d$ dimensions the cost of tomography (measurements and classical storage) become $C^{\mathcal{O}(dk)}$ while that of reconstructing the unitaries is $\mathcal{O}(k C^d T_e)$, where $T_e$ is the cost of computing one entry of $A[m]$.

In the context of hadron physics, the Hilbert space dimensionality, the breaking of perturbative expansions and the Monte Carlo sign problems of LQCD are some of the reasons that explain the slow progress towards the characterisation of the QCD phase diagram and in particular of the nature of the phase transition between deconfined quark matter and confined hadronic or nuclear media. Such a phase transition is expected to occur at the core of massive Neutron Stars\cite{Baym:2017whm}. Thus one can try to derive a suitable EoS by approaching the phase transition from the quark-matter side and to compare it with the astrophysical constraints. Such a system should be described by the Grand-Canonical ensemble, in which the fixed thermodynamic variables are the temperature $T$, the volume $V$ and the chemical potential $\mu$, with partition function:
\begin{equation}
    Z(\beta,\mu) = \text{Tr}\,e^{-\beta (H-\mu N)}\ ,
    \label{Grand-Canonical}
\end{equation}
in the low temperature limit this reduces to a ground-state-search problem, for which algorithms such as VQE could be used. The tricky part is then to construct suitable ansätze, which will necessarily depend on the concrete realisation of QCD; a blind variational search will be probably plagued with barren plateaus. 

Our group is exploring an approach to calculate the ground-state energy from the particle encoding in section~\ref{sec:particleregisters} within Weyl-gauged QCD, with a simple Hamiltonian but the need to enforced Gauss' law independently, see subsection \ref{subsec:Weyl}. 
To be more specific, we divide the system into small and independent cells with periodic boundary conditions in which the final aim is to minimize the cost function
\begin{equation}
    E(\vec{\theta};\mu_f,\nu) = \left\langle\psi(\vec{\theta})\right|\hat{H} + \mu_f \hat{N}_f+ \nu \,\hat{\mathcal{G}}^2\left|\psi(\vec{\theta})\right\rangle\ ,
\end{equation}
on a quantum memory with a few dozens of particles. $\hat{H}$ is the QCD Hamiltonian in the Weyl-gauge, $\hat{\mathcal{G}}^2$ is the Gauss operator squared\footnote{ With $\mathcal{G}$ a self-adjoint operator, $\left\langle\psi\right|\mathcal{G}^2\left|\psi\right\rangle=\sum_\phi\left\langle\psi\right|\mathcal{G}\left|\phi\right\rangle\left\langle\phi\right|\mathcal{G}^\dagger\left|\psi\right\rangle=\sum_\phi|\left\langle\psi\right|\mathcal{G}\left|\phi\right\rangle|^2\geq 0$: its minimization corresponds to $\langle \phi | \mathcal{G}=0$ which ensures the gauge choice by enforcing Gauss's law.} and summed over positions and colours, $\hat{N}_f$ is the number of $f$-flavoured fermions, $\vec{\theta}$ are the variational parameters, $\mu_f$ are the chemical potentials and $\nu$ is a Lagrange multiplier. The EoS can then be constructed by analysing the change in energy density with the volume of these cells and employing standard statistical physics. To give an example on the use of the encoding of sec. \ref{sec:particleregisters} to measure expectation values, consider the splitting part of the three gluon vertex in Weyl gauge, 
\begin{equation}
    H_{g21} = g\int\left[p,k,q\right]f^{abc}\left(p^j\delta^i_l + k^l\delta^i_j+q^j\delta^i_l\right)\left(i\,a^\dagger_{pia}a^\dagger_{kjb}a_{qlc}+h.c. \right)\tilde{\delta}\left(k+q-p\right)\ ,\end{equation}
with a similar term for the Gauss operator squared
\begin{align}
\mathcal{G}_{g21}^{2}=2g\int\left[kpq\right]f^{abc}&\left(E_{k}E_{p}p^{i}\delta_{lj}+E_{q}E_{p}q^{l}\delta_{ji}\right. \nonumber\\
&  \left.+E_{q}E_{p}p^{i}\delta_{lj}\right)\left(ia_{pia}^{\dagger}a_{kjb}^{\dagger}a_{qlc}+h.c.\right)\ \tilde{\delta}\left(p+k-q\right)\ ,\end{align}
with $\int\left[k\right]=\int\frac{d^{3}k}{\left(2\pi\right)^{3/2}\sqrt{2 E_{k}}}$ and $\tilde{\delta}\left(p\right)=\left(2\pi\right)^{3}\delta^{3}\left(p\right)$.
If the variational state is properly symmetrized and the memory is filled in order, the product of Eqs.~(\ref{def:nReg-bosoncreation},\ref{def:nReg-bosonannihilator}) simplify to:
\begin{align}
\bra{\psi}a_{\rho}^{\dagger}a_{\eta}^{\dagger}a_{\omega}+h.c.\ket{\psi}=\bra{\psi}  \left\{ \sum_{j=1}^{N_{f}}j\sqrt{j+1}\ \projector{1}{0}\otimes\right.&\left(\ketbrac{1}{0}\otimes\qset{\rho}\right)_{j+1}\otimes\nonumber \\
&\left.\left(\ketbrac{1}{1}\otimes\qset{\eta}\qscrap{\omega}\right)_{j}\otimes\projector{1}{1}+h.c.\right\} \ket{\psi}\ ,
\end{align}
where $\rho,\eta,\omega$ are compact indices for the gluon quantum numbers. This expression provides a LCU decomposition of the Fock operator product in terms of unitary matrices, since the projectors $\mathbb{P}$, controls $\mathfrak{C}$ and set $\set{}$ and scrap $\scrap{}$ operators can all be written in terms of products of Pauli matrices.
\section{Progress towards Fragmentation Functions} \label{sec:fragmentation}

As described in subsection~\ref{subsec:frag}, the computation of fragmentation functions
is an obvious target for quantum computing investigations. Our group has recently produced a limited computation~\cite{Galvez-Viruet:2025rmy} 
of the fragmentation function $D_c^{J/\psi}$ for a charm-quark initiated jet to fragment into a $J/\psi$ meson. 

For this, we have completely recalculated the Hamiltonian of Quantum Chromodynamics in Light Front Quantisation and in Light Front Gauge~\cite{Brodsky:1997de}. We adopted a jet initiated by a charm quark with a large starting longitudinal light-front momentum $p_j^+=\Lambda$ that sets the maximum of our longitudinal grid, so that the emitted $J/\psi$ carries a fraction $z$ given as $p_h^+=z p_j^+$. 
We employed three qubits (eight longitudinal momentum states) per particle, and neglected all transverse momenta (in principle not a terrible approximation in jet fragmentation where $p_\perp \ll p_+$). To this we need to add colour, flavour if the parton is a quark or antiquark, and spin, to total 7--8 qubits per parton.  

We employed the encoding described in Section~\ref{sec:particleregisters} to store up to four particles (the initiating $c$-quark plus an additional gluon and an additional $c\bar{c}$ pair).
This, plus a few additional ancillary qubits, tallies up to 30 qubits which make a large Hilbert space with basis cardinal $2^{30}\simeq 10^9$: this is near the limit of possible classical simulators, but if the $O(100)$ qubits of the IBM chips would in the future allow a large number of entangling gates to be sequentially executed, we could soon see large particle-bases and very realistic computations of fragmentation functions.

A few calculational details are in order.
First, to extract the Fragmentation Function from Eq.~(\ref{FuncionFragmentacion})
we make an ansatz $J/\psi$ longitudinal distribution
\begin{equation}
|J/\Psi\rangle =  \sum \delta_{c_qc_{\bar{q}}}  \frac{\chi_0(x) \,\vec{\sigma}_{ij}}{\sqrt{x \,(z-x)}}
\left|x\ i\,c_q ,(z-x)\ j\,c_{\bar{q}}\right\rangle\ ,
\end{equation}
where the longitudinal momentum distribution inside the meson is 
\begin{equation} \label{Cisneroschi}
\chi_0(x) = \frac{1}{\sqrt{C}} \  x^{\beta/2} (z - x)^{\alpha/2}\ ,
\end{equation}
suppressed at both ends $x=0$ (the valence charm quark carries a negligible fraction of the $J/\psi$ longitudinal momentum, thus also of the original jet) and $x=z$ (the valence charm constituent carries the entire $J/\psi$ $p^+_{J/\psi}$ and therefore the entire $z$ fraction which was fragmented of the jet.

\begin{figure}
    \centering
    \includegraphics[width=0.6\linewidth]{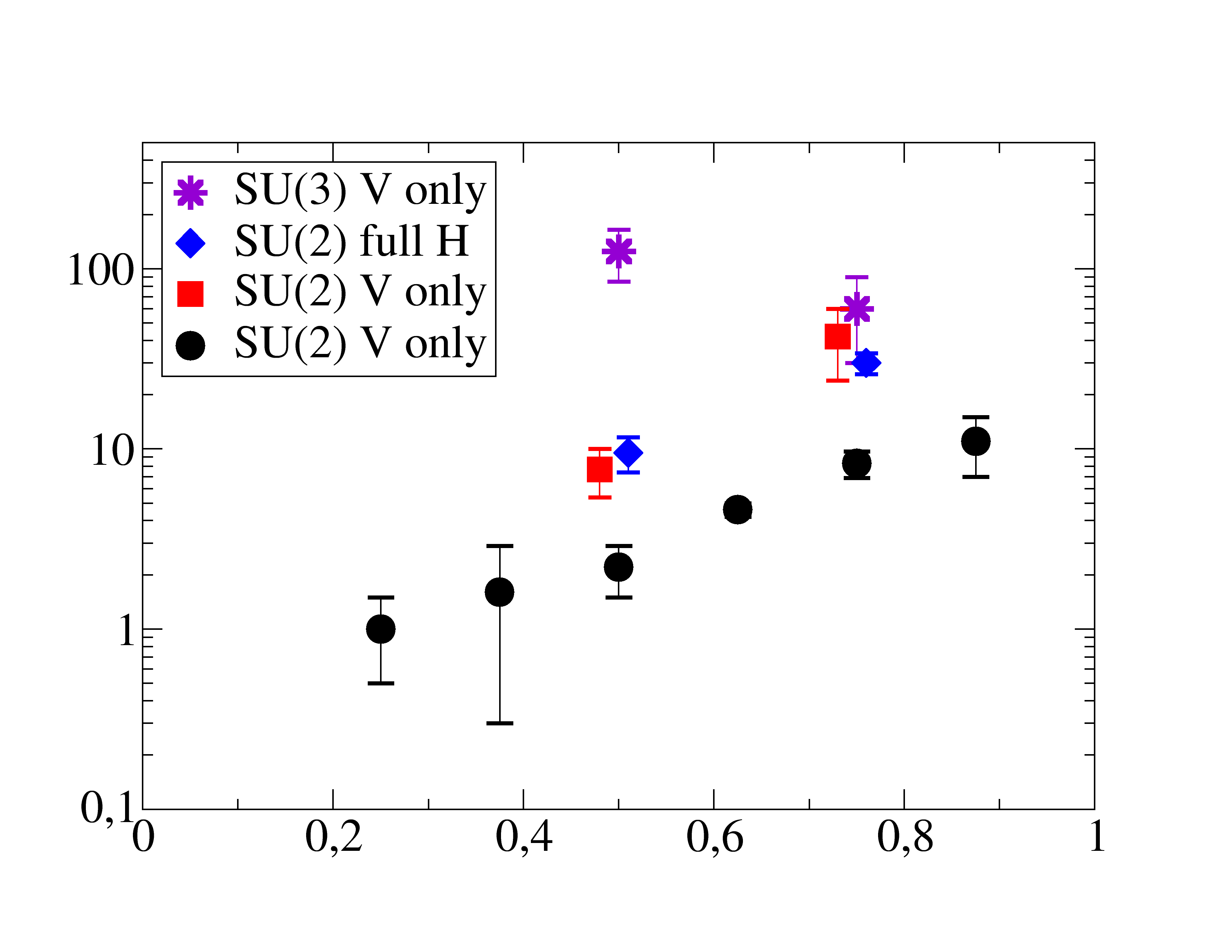}
    \caption{Fragmentation function $D_c^{J/\psi}$ extracted from a classical simulation of a quantum computer with order 30 qubits. An $SU(3)$ calculation (stars) was run with only part of the interaction vertices (splitting, absorbing and pair creation) and one can see, from the equivalent $SU(2)$ computation at equal $\alpha_s$ and conditions (squares) the dependence on the colour factor. The entire Hamiltonian was run only in the $SU(2)$ case (diamonds) but the difference is seen to be small. These three computations had a meager $N=4$ partition of the longitudinal momentum; an additional $N=8$ computation (circles) is shown to indicate the sensitivity thereto.}
    \label{fig:fragJpsi}
\end{figure}

The steps of the algorithm are as follows:
\begin{compactenum}[a)]
\item Prepare the memory containing only the $c$ quark, at the maximum momentum of the grid, as the jet primer.
\item Evolve this protojet with the light-front time evolution operator employing a Trotter expansion to separate the various terms of the Hamiltonian, $U( \Delta t)=e^{-iT \Delta t} \prod_j  e ^ {-iH_{Ij} \Delta t} + O((\Delta t)^2)$.
\item In that evolution, because $H_I$ contains particle-number-changing operators such as $c^\dagger c a^\dagger$, the particle number evolves and the state becomes a superposition of various numbers of quarks and gluons. 
\item Given enough memory, the number of particles would become large (``saturation regime'') and entropy would be maximised once the probability would be spread among the many degrees of freedom. In our simulation, with at most four particles, the unitary operator can start cycling back. We compute $S$ and look for a plateau.
\item At that plateau we consider, for the time being, that $x^+\to \infty$ and extract the fragmentation function. 
\end{compactenum}
The outcome of an example computation is shown in Fig.~\ref{fig:fragJpsi}. The grid cannot be very tight (due to the small number of qubits which can be simulated), the maximum number of particles is 4 (up to two quarks, an antiquark and a gluon) and either the group was shortened to $SU(2)$  or not all terms of the Hamiltonian (particularly seagulls and fork interactions germane to the light-front quantisation) were employed in all calculations; but in spite of these limitations, necessary to obtain finite running times on a standard computer cluster,  we believe that we have demonstrated the feasibility of extracting fragmentation functions from upcoming quantum computers with a beginning-to-end algorithm that serves as a demonstrator.

\paragraph{ Parton Distribution Functions (pdfs),} the distributions parametrising the inverse problem --that of extracting a parton constituent from a hadron instead of the fragmentation of a hadron from a jet--
have also been an object of recent scrutiny. For example, in the Nambu-Jona-Lasinio model in 1+1 dimension~\cite{Kang:2025xpz}, the Schwinger model~\cite{Banuls:2025wiq,Chen:2025zeh}
(see subsection~\ref{subsec:Schwinger} below for a brief discussion thereof).

Typically, the resulting distribution functions in this type of constituent-like models are valence-like, without the characteristic low-$x$ growth revealed by HERA-age experiments~\cite{ZEUS:1993ppj}. 
One would expect that the ability of the quantum computer to one day manage large numbers of particles would allow to see the low--$x$ growth of parton distributions driven by multigluon emission. For the time being, the regime that quantum computers can access is the same that lattice gauge theory probes via moments or the LaMEFT (Large-Momentum Effective Field Theory), so that we do not see it as an immediate priority.

\color{black}
\newpage
\section{Progress towards time evolution in hadron physics} \label{sec:tevol}

\begin{figure}
    \centering
    \includegraphics[width=0.6\linewidth,trim=0 0 10 10,clip]{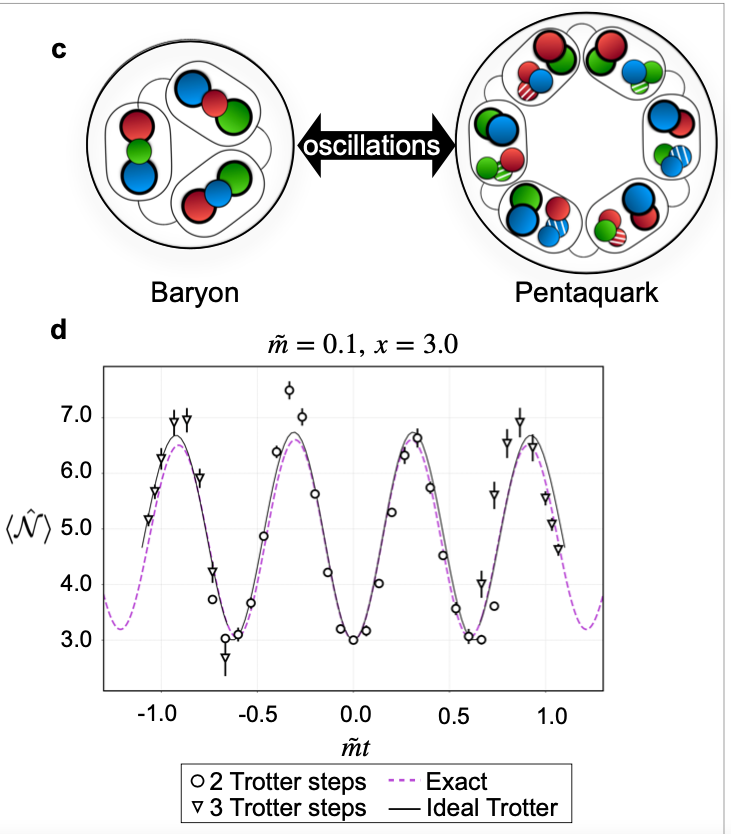}
    \caption{
    Example of time-evolved computation of a baryon, showing the oscillation between the minimum 3--quark component and Fock--space sectors with a larger number of particles, {\it e.g.} pentaquark--like configurations.
    {\it Reprinted\cite{Atas:2022dqm} under the terms of the Creative Commons 4.0 License,} {\tt https://creativecommons.org/licenses/by/4.0/}}
    \label{fig:TrotterPentaquark}
\end{figure}

\begin{figure}
    \centering
\includegraphics[width=0.65\linewidth]{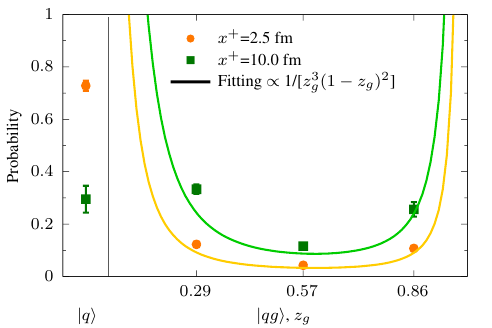}
    \caption{Example of light-front-time evolved computation starting with a quark in a jet and measuring its splitting (or not) to a quark-gluon pair at later $x^+$ of 2.5 and 10 fm, respectively, and in the second case, taking three measurements of the longitudinal momentum fraction.
    {\it Reprinted\cite{Barata:2023clv} under the terms of the Creative Commons 4.0 License,} {\tt https://creativecommons.org/licenses/by/4.0/}}
    \label{fig:splittingJet}
\end{figure}

A strength of the quantum--computing formulation of Chromodynamics is the possibility of following up in Minkowski space the time evolution of a hadron system, be it canonical time $t$ or light--front time $x^+$. Two characteristic examples are shown in figures~\ref{fig:TrotterPentaquark} and~\ref{fig:splittingJet}.

The first one~\cite{Atas:2022dqm} is an $SU(3)$ gauge theory with the canonically quantised lattice Hamiltonian already described (section~\ref{sec:lattice}).
An interesting method employed to mitigate quantum computer error consisted in computing a physics run evolving forward for $N_T$ Trotter time steps; and then calculate again, evolving forward for $N_T/2$ steps and backwards for another $N_T/2$, with the aim of reconstructing the initial state, whose comparison with the original preparation allows to characterise the computer error.
The number of time steps $N_T$ was of order 4 to 8 depending on the machine employed.
The evolution of both the quark + antiquark particle number (shown in the figure) as well as that of the gauge field were tracked.

Another example of this type of calculation, again for QCD in 1+1 dimensions, can be found in \cite{Farrell:2022wyt}, the Kogut-Sussking Hamiltonian of Eq.~\ref{KSH} is formulated in the axial gauge $A^x=0$, allowing for the gauge degrees of freedom to be transformed into long-range interactions between quarks. This generates ``colour edge'' states at the end of the lattice which are artificially ``pulled up" in energy adding to the Hamiltonian a term that depends on the sum of colour charges. The problem is formulated on a lattice with quarks and antiquarks of two flavours, requiring $3\times 2\times 2 = 12$ qubits per lattice site. The low-energy spectrum is solved classically for several values of the coupling constant $g$ and with a quantum annealer, a special device to solve variational problems. Later the authors use the Jordan-Wigner transform to write the Hamiltonian operator in terms of Pauli operators, and develop efficient circuits for the simulation of real-time evolution. The results, which closely resemble exact evolution, are obtained with aid of error-mitigation techniques. Closely related is\cite{Farrell:2022vyh}, in which the authors add to the Hamiltonian an effective four lepton interaction to generate baryon $\beta$ decays. The simulation is run on a $20$ trapped-ion quantum computer\cite{Quantinuum} filtering the results that are obvious errors (post-selection) without error mitigation, obtaining results at percent level.

The final example corresponds to a Trotter--expanded time evolution of the light-front $P^-_{\rm QCD}$ Hamiltonian (see subsection~\ref{subsec:lightfront} for our take on it). A quark--initiated jet~\cite{Barata:2023clv} is allowed to interact, in the eikonal approximation, with a modeled gluon medium. Several observables were extracted from the quantum simulation, aiming at the energy loss of the jet, but we highlight in figure~\ref{fig:splittingJet}
the phenomenon analogous to that in the earlier figure~\ref{fig:TrotterPentaquark}, namely the splitting of the quark in the jet to a quark-gluon pair, showing both the probability of finding a quark alone and the probability density of having a quark and a gluon with certain momentum fractions.

The figure shows quite clearly that the method is ready for actual physics work as soon as the number of particles can be scaled with larger memory banks in upcoming quantum computers.

These computations demonstrates that the multiparticle Fock space nature of the theory can now be explored (this has, historically, been a very nontrivial endeavor with many--body methods~\cite{General:2007bk}). We look forward to the extension of this type of work to the prediction of actual experimental observables.
\subsection{Jet broadening}

We now briefly discuss a calculation~\cite{Barata:2022wim} within the light-front Hamiltonian formalism showcasing three-dimensional
jet evolution in a medium, which demonstrates transverse momentum broadening.
The interest in having a quantum computer assess this problem is that perturbative calculations exist but are supposed to be valid at large temperatures and virtualities of the jet's leading parton. However, many observations of jets do not satisfy those stringent conditions and perturbation theory is more questionable, with phenomenological approaches~\cite{Hidalgo-Duque:2013rta} suggesting larger values of the quenching parameter. Once more, while there are lattice gauge theory computations~\cite{Kumar:2019aop} attempting to extract $\hat{q}$, the jet quenching parameter, this is a quantity defined along a lightlike path, and in an Euclidean computation some form of analytical extrapolation on noisy data is needed, entailing large uncertainties. In fact, for temperatures below 300 MeV, the data there reported is not very meaningful. So this can be another project where quantum computers make headway.

The Light-front Hamiltonian for the parton probe driving the jet is split~\cite{Barata:2022wim} as
$P^- = \frac{{\bf p}_\perp^2}{2p^+}+ g T^a A^{a\ -} $ (because other components of $A$ are suppressed by powers of $p^+$, the large scale). The dominant component is then taken to satisfy the classical Yang-Mills equations with a local source.
The computed physical quantity, the  quenching parameter $\hat{q}$, is defined from the broadening of the jet (per unit length $L$) in transverse momentum, 
\begin{equation}
\hat{q} = \frac{\langle {\bf p_\perp^2(L)} \rangle - \langle {\bf p_\perp^2(0)}   \rangle}{L}\ .
\end{equation}
The algorithms employed are akin to those described above, so we refer the reader to the original publication~\cite{Barata:2022wim}.
Due to memory restrictions with a small number of qubits, the gauge groups chosen were $U(1)$ and $SU(2)$. 
For the latter, closer to QCD, the quenching parameter is obtained to be about $\hat{q}\simeq 0.24$GeV$^3$ at a scale $Q_s^2\simeq 5 $GeV$^2$ and with a gluon masslike parameter $m_g\simeq 0.1$ GeV (which must give an indication of the scale of the temperature if the medium is thought of as thermal).

Jet energy loss is also discussed in a  1+1 Schwinger model with a wave packet plowing through a static medium in subsec.~~(\ref{subsec:jetS}) below. Because of the one-dimensional setup, particles can separate only in the longitudinal direction instead of opening a cone with transverse momentum, so many features of jet physics are absent.  
Other approaches to study jet propagation include the use of the open quantum system formalism with a scalar field-theory mock of the QGP medium~\cite{DeJong:2020riy}.

\color{black}
\section{Applications from encoding phenomenological Hamiltonians} 
Quantum computers are expected to efficiently simulate hadron dynamics \cite{Jordan:2012xnu}. However, since a full treatment of the problem is not feasible in the present NISQ era, current research focuses on developing algorithms within simplified theories. We now review some recent results on fragmentation functions in the NJL model, jets, energy loss and entropy production for ``hadrons'' in the Schwinger model and spectroscopy of phenomenological models.

\subsection{Fragmentation in the NJL model and the Schwinger model} \label{subsec:Schwinger}

\begin{figure}
    \centering
    \includegraphics[width=0.65\linewidth]{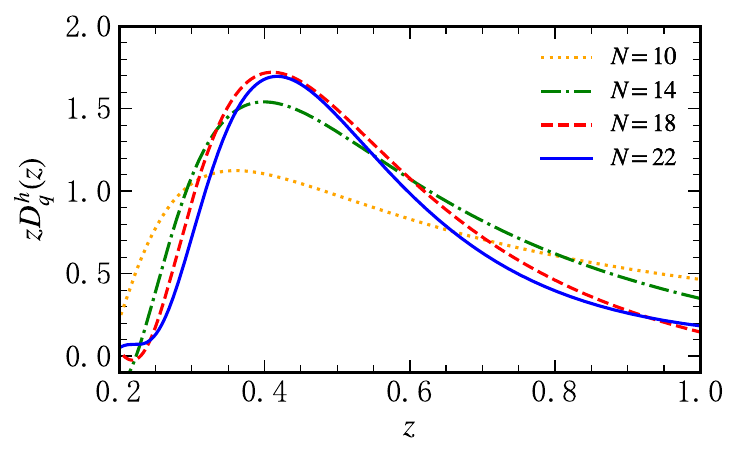}
    \caption{Fragmentation function for a one-flavour meson in 1+1 dimensions computed in the Nambu-Jona-Lasinio model, showing the convergence as a function of the number $N$ of qubits . 
    {\it Reprinted~\cite{Li:2024nod} under the terms of the Creative Commons 4.0 License,} {\tt https://creativecommons.org/licenses/by/4.0/} }
    \label{fig:FFNJL}
\end{figure}

A computation of a hadron's fragmentation function  in a spacetime lattice, with staggered fermions implementing the Nambu-Jona-Lasinio quark model given by the Lagrangian (for one flavour and in 1+1 dimension, that is, ignoring the transverse dimension of hadrons)
\begin{equation}
\mathcal{L} = \bar{\psi} (i\gamma^\mu \partial_\mu -m)\psi + g (\bar{\psi}\psi)^2\ ,
\end{equation}
is reprinted in Figure~\ref{fig:FFNJL}. This is a computation on a classical simulator advancing what could be done on a quantum computer with sufficient circuit depth (the modest number of qubits, up to 22 here, is already available as described in section~\ref{sec:hardware}).

The circuit depth (proportional to the number of gates which need to be executed) is polynomial~\cite{Li:2024nod}, with a cost $\propto O(N^3/\epsilon)$ with $\epsilon$ the required accuracy level and $N$ the number of qubits.

There are numerous approximations in this computation: the gluons have been suppressed and the only interaction is the contact vertex of the NJL model. 
The inclusive $X_{\text{out}}$ state accompanying the fragmented hadron in $| h,X_{\text{out}}\rangle 
\langle h,X_{\text{out}} |$ at Eq.~(\ref{FuncionFragmentacion}) is reduced to the vacuum $|\Omega\rangle$ and one--particle $| \Psi_1 \rangle$ states. The meson mass is taken to be 1.5 times the quark mass instead of being computed, consistently, from the same Hamiltonian, etc. 
Still, this demonstration of principle confirms that a working quantum computer will open new calculational possibilities which remain, to date, untapped. 

There is also a recent extant calculation of the equivalent to fragmentation functions in the Schwinger model~\cite{Grieninger:2024axp}
(as well as one of quasi parton distributions~\cite{Grieninger:2024cdl}). The technique involves computing ``quasi fragmentation functions'',  
a related quantity where the two field insertions are not separated along the light-front, but are successively approaching it by means of boosts.
In terms of the rapidity function associated to a velocity $v\in[0,1)$,
\begin{equation}
\eta = \frac{1}{2} \log \left(
\frac{1+v}{1-v}
\right)\ ,
\end{equation}
the Lorentz dilatation factor $\gamma(v)$, the boost operator $K$ and the Hamiltonian $H$, the authors propose
\begin{equation}
    e^{-i\eta K} e^{-iHt}\ \overline{\psi}(0,\pm x^3)
    e^{iHt} e^{i\eta K} = \overline{\psi}(
    -\gamma (t\pm vx^3) , \gamma (vt\pm x^3 )
    e^{-\frac{1}{2}\eta\gamma^5})\ ,
\end{equation}
so that, from evolving a field at spacelike separation $\pm x^3$ from the origin, which is accessible on an ordinary lattice formulation,  and letting $v\to 1$, one can hope to extract quantities along the light front.

\subsection{Jet energy loss and entropy production}
\label{subsec:jetS}

Another recent interesting application of quantum computing to simple Hamiltonians inspired in QCD is the computation of the interaction of a jet with a medium~\cite{Barata:2025hgx} (remember that jet quenching is an ubiquitous probes of the medium formed in ultrarrelativistic heavy-ion collisions).
The model is the $A_0=0$ Weyl--gauge 1+1 dimensional Schwinger model, with Hamiltonian
\begin{equation}
    H= \int dx \left( \frac{E^2(x)}{2} +
    \overline{\psi}(x)
    (\gamma^x(-i\partial_x + g A_x(x) )+m\mathbf{1})
    \psi(x)
    \right)\ .  \label{Schwinger1plus1}
\end{equation}
(The model has also been deployed to estimate spin correlations~\cite{Barata:2023jgd})

\begin{figure}
    \centering
\hspace{-0.5cm}\includegraphics[width=1.1\linewidth]{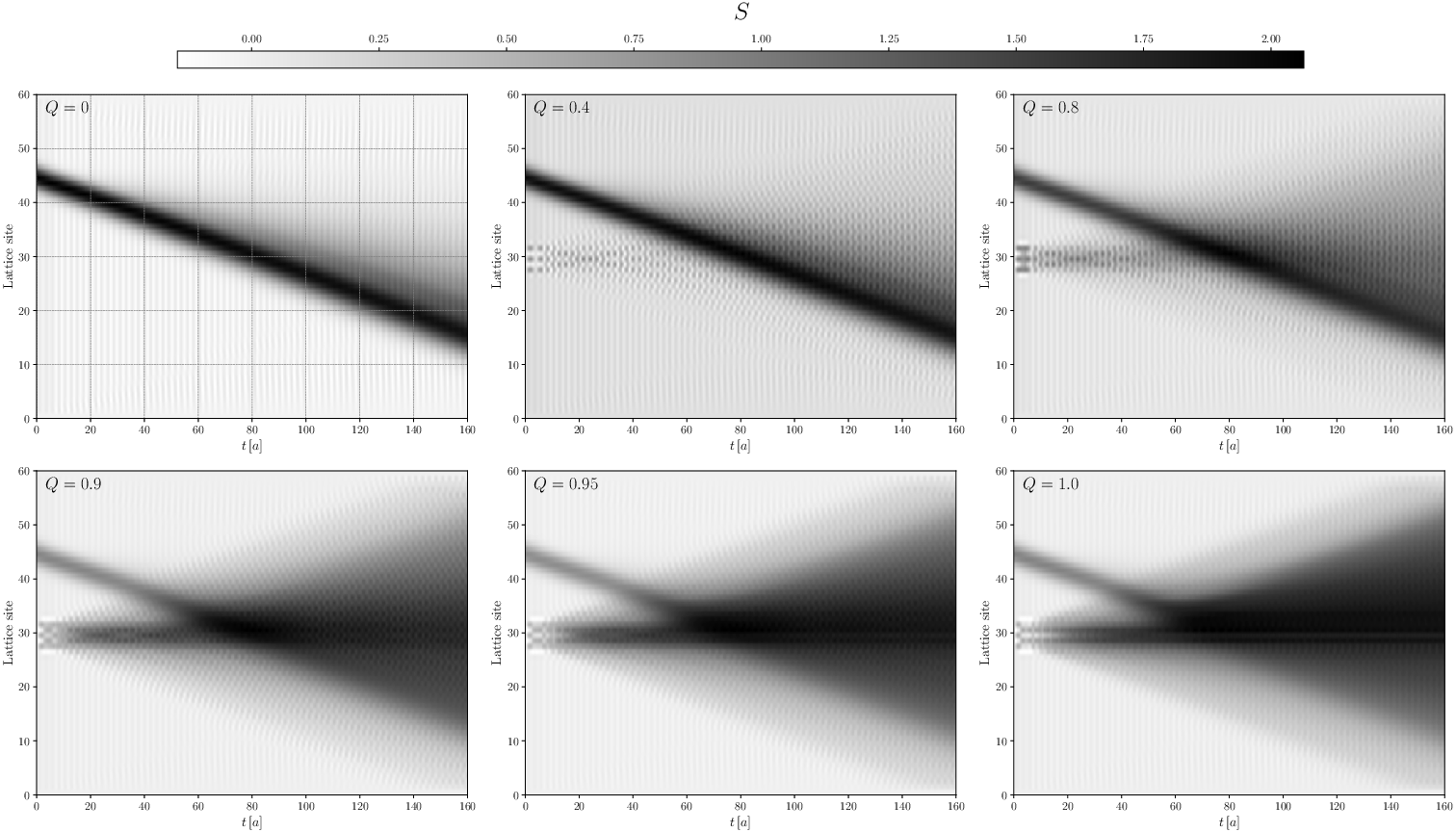}
    \caption{Heat--map type plots of the entropy produced by a jet (coming in from the top left in each plot) scattering in a slice of a medium (centre of the figure) for different values of the total charge of the jet in the Schwinger model of Eq.~(\ref{Schwinger1plus1})
    {\it Reprinted\cite{Barata:2025hgx} under the terms of the Creative Commons 4.0 License,}  
    {\tt https://creativecommons.org/licenses/by/4.0/}}
    \label{fig:Jetentropy}
\end{figure}

Employing a spacetime lattice setup with staggered fermions, the time evolution of a system composed of a ``jet'' (fast charge of the Schwinger model) traversing a medium was calculated, in particular the energy loss of the jet and the distribution of entropy, which we reproduce in figure~\ref{fig:Jetentropy}.

The computation was carried out with an efficient classical simulator (tensor network) of the behaviour that a quantum computer would display.
From left to right and top to bottom the charge coupling the jet to the medium increases, and one sees how an increasing amount of entropy is generated and dispersed.

\subsection{State preparation and energy loss in charged medium}
Research on state preparation concerns the construction of appropriate ansätze and the minimisation of objective functions such as the expectation value of the energy or the infidelity with respect to a desired target state. Theoretical and practical inputs about the process are based on the preparation and evolution within simplified models, such as the 1+1 D Schwinger model\cite{Farrell:2024fit}, in which the ``hadrons'' are $e^+-e^-$ bound states, exemplified with the use of the $112$ qubits quantum-computer \textit{ibm\_torino}\cite{IBMquantum:2025}. The vacuum of the theory is prepared first, and subsequently a ``hadron'' on top of it, keeping track of the evolution of the chiral condensate expectation value as the system evolves in time for up to $14$ Trotter steps. 
The state-preparations procedure is named \textit{SC-ADAPT-VQE} and consists on repeatedly tuning selected unitary transformation with classical circuits for increasing lattice lengths until convergence, finally extrapolating these transformations to the real length on the quantum device. It is found that classical simulators can manage the preparation for quite large lengths as the correlations between physical sites remain small, but run into trouble for time evolution.

The problem has to be somehow further simplified to reduce the scaling of the interactions from quadratic in the lattice length $\mathcal{O}(L^2)$ to linear $\mathcal{O}(\lambda L)$: arguing that correlations between charges at different spatial sites fall-off exponentially, the interactions are cut to qubits separate at most by $\lambda$ sites. These approximations are combined with error mitigation techniques to extract meaningful results from the noisy raw data obtained from the devices.

The preparation and time-evolution circuits of\cite{Farrell:2024fit} are combined with the presence of static charges to mimic, via classical simulators, the evolution of heavy particles in dense media \cite{Roland:2025}. Lattice effects are probed through the evolution of an initial bound state, where the light dynamical degrees of freedom couple to a heavy charge. Due to the lattice dispersion relations, once a critical velocity 
$v^{*}$, less than the speed of light, is exceeded, the light degrees of freedom decouple from the charge and excite the lattice in a process resembling hadronisation. To suppress such effects, the velocity of the moving charge is kept small as it traverses the lattice populated with other static charges. Within this setup, the energy loss and entanglement are computed, allowing lattice discretisation effects to be isolated. Entanglement is found to be more sensitive to lattice artifacts than other classical observables. In addition, quantum circuits and associated computational costs are discussed.

\subsection{Real Time Scattering}

Another application where there is interest is in tracking real--time scattering among hadrons. This is not possible in the lattice, a Euclidean formulation, so it has been thought of~\cite{Papaefstathiou:2024zsu} as a possible application for quantum computers.

We are a bit less enthusiastic than other authors about this real-time scattering as a flagship application in hadron physics. Indeed, lattice gauge theory is already extracting scattering phase shifts and inelasticities, pole positions, etc. through analysis of the Lüscher formula (via the relation of the density of states at a given energy for fixed angular momentum and the phase shift of the scattering, as lattice methods are very competent at identifying spectral energies). We see the interest of real-time scattering to be more intense in lower-energy fields, such as biochemistry, where there is hope to experimentally follow individual reactants in time. For hadron physics, and for the time being, this is more of a curiosity. This is of course opinionable, so we leave here a mention of the effort and refer the reader to the literature.

\color{black}
\section{Outlook}

There are dedicated reviews on Quantum Computing which cover topics in High-Energy Physics~\cite{DiMeglio:2023nsa} and Nuclear Physics~\cite{Garcia-Ramos:2023rtd}, but we feel that
there is a need to point out the important niche applications in Hadron Physics, straddling both fields, sharing from the large scales of the first but also from the nonperturbative problems of the second. We have striven to give, in a few brushstrokes, a colourful picture of  some interesting
problems which can be addressed when quantum computers become a more practical tool, without precluding that several other interesting applications may be devised in the field.

The underlying theory of hadrons is Quantum Chromodynamics (QCD), a rich framework that has been extensively explored over the past five decades through a wide variety of theoretical and numerical approaches. Quantum computers are, in this sense, promising instruments that offer the opportunity to revisit longstanding problems from new perspectives—as illustrated by studies on gauge fixing, fragmentation, and spectroscopy, to mention only a few examples. Alongside these efforts, many exploratory works have appeared in recent years, seeking to demonstrate the potential of these emerging machines through tangible examples while pushing existing hardware to its limits. Nevertheless, the technology remains in its infancy, and its ultimate impact on fundamental research is still difficult to foresee.

This entails that several extant works fall back onto topical phenomenological models addressing one or another aspect of the full theory. As hardware advances, the advantage of directly working with discretisations and truncations of QCD lies in the possibility of controlling systematic uncertainties through well-defined computational parameters—a feature that also applies to quantum simulations.

Finally, one crucial aspect that quantum computing practitioners entering the field of quantum field theory simulations should bear in mind is the renormalisation, or flow, of the Hamiltonian parameters when the scale of the problem changes (and hence the discretisation), whether in momentum space or particle-number space. The following paragraph is devoted to this issue.

\paragraph{Renormalisation Group evolution}
\label{sec:RG}
We have not dedicated space in this brief appraisal to the renormalisation of the QCD Hamiltonian, but a comment is in order here, since, unlike the quantum mechanical theories often transcribed for a quantum computer, Chromodynamics will require serious work to handle its renormalisation intricacies. A promising line of research is that of the Renormalisation Group Procedure for Effective Particles (RGPEP)~\cite{Glazek:2016ifz} recently deployed for a Yukawa Hamiltonian~\cite{Gustin:2025ogl}.
The traditional idea of continuously transforming the Hamiltonian via a spectrum-conserving unitary rotation parametrised by a control variable $s$ (for example, a scale)
\begin{equation}
H(s) = U(s) H(0) U^\dagger(s)\ ,
\end{equation}
is there deployed with a choice of $U$ such that the Hamiltonian evolves with a double commutator,
\begin{equation}
    \frac{dH(s)}{ds} = [[H_{\rm free},H(s)],H(s)]\ .
\end{equation}
New counterterms are then added to the Hamiltonian and the eigenvalues are extracted. This, or other RGE techniques~\cite{Robertson:1998va,Glazek:2015nil,Leder:2011uoc} will need to be automated into the workflow for QCD on quantum computers, which is currently best described as simply cutoff (regulated).

The RGPEP is most often formulated using the Light Front, and the Hamiltonian requires renormalisation. The infinite number of degrees of freedom requires truncation; to study the sensitivity thereto, calls for a renormalisation-group method to be applied. The RGPEP enables the construction of effective Hamiltonians through a similarity transformation that depends on a scale parameter. The eigenvector of the effective Hamiltonian contains a small number of non-negligible Fock components,  thereby reducing the complexity of the description. Such an approach appears promising for reducing the number of degrees of freedom without truncating terms that encode essential dynamics. Simulations with effective Hamiltonians have been performed for a Yukawa theory within RGPEP, showing that the computational cost to block encode the renormalised Hamiltonian is comparable to block encoding the bare Hamiltonian~\cite{Serafin:2025ouo,Gustin:2025ogl}.

\paragraph{Conclusion}
We put an end to this brief review on a positive note. In spite of all the hurdles which need to be overcome so quantum hardware has real impact in hadron physics, recent progress has been substantial and it might well be that within a decade there are machines which run realistic calculations. Meanwhile, we should strive to continue advancing algorithms and identifying problems which can be profitably addressed in those early machines.

\section*{Acknowledgements}
Work partially supported by grants 
PID2023-147072NB-I00; 
PID2022-137003NB-I00 
of the Spanish MCIN/AEI /10.13039/501100011033/; grant FPU21/04180 of the Spanish Ministry of Universities.

\section*{ORCID}
\noindent J. J. G\'alvez-Viruet   - \url{https://orcid.org/0000-0003-3725-7267}\\
\noindent Felipe J. Llanes-Estrada  - \url{https://orcid.org/0000-0002-2565-4516}\\
\noindent Mar\'{\i}a G\'omez-Rocha  - \url{https://orcid.org/0000-0002-9513-5797}

%


\end{document}